\documentclass[a4paper,11pt]{article}
\pdfoutput=1 
\usepackage{graphicx}  
\usepackage{dcolumn}   
\usepackage{bm,relsize}        
\usepackage{amssymb, amsmath}
\usepackage{textcomp}
\usepackage{wasysym}
\usepackage{slashed}
\usepackage{caption, subcaption}
\usepackage{multirow}
\usepackage[para,online,flushleft]{threeparttable}

\usepackage{lipsum, color}
\usepackage[usenames,dvipsnames,svgnames]{xcolor}
\usepackage{braket}
\usepackage{jheppub} 
\usepackage[dvipsnames]{xcolor}
\usepackage{tabularx}
\usepackage{soul}

\usepackage{microtype}   
\usepackage{amsmath}    
  \usepackage{amsthm}    
  \usepackage{amssymb}  
\usepackage{siunitx}    
\usepackage{booktabs}    
\usepackage[capitalize]{cleveref}  
  
  \crefname{section}{Sec.}{Secs.}
  \crefname{appendix}{App.}{Apps.}

\usepackage[shortlabels,inline]{enumitem} 
\usepackage{mathtools}
\usepackage{subdepth}  

\usepackage{graphicx,multirow}
\usepackage{float} 
\usepackage[T1]{fontenc}
\usepackage{physics}
\usepackage{bbold}
\usepackage{rotating}
\setlist[enumerate,2]{leftmargin=0.45em}
\usepackage{comment}
\usepackage{xspace}

\usepackage{tikz}
\usetikzlibrary{arrows.meta, positioning, shapes.geometric, calc, fit, backgrounds}

\DeclareFontFamily{U}{mathx}{\hyphenchar\font45}
\DeclareFontShape{U}{mathx}{m}{n}{<-> mathx10}{}
\DeclareSymbolFont{mathx}{U}{mathx}{m}{n}
\DeclareMathAccent{\widebar}{0}{mathx}{"73}

\newcommand{\beq}{\begin{equation}}
\newcommand{\eeq}{\end{equation}}

\newcommand{\lhp}[1][]{\ensuremath{{\mathcal{P}_{#1}}}\xspace}
\newcommand{\llh}[1][\text{NF}]{{\ensuremath{\mathcal{L}_{#1}}}\xspace}
\newcommand{\bz}{\ensuremath{\boldsymbol{z}}\xspace}
\newcommand{\bc}{\ensuremath{\boldsymbol{c}}\xspace}
\newcommand{\bx}{\ensuremath{\boldsymbol{x}}\xspace}
\newcommand{\mmp}[1]{\,#1}
\newcommand{\btheta}{\ensuremath{\boldsymbol{\theta}}\xspace}
\newcommand{\Dd}{\Delta\delta}
\newcommand{\KK}{\bar{K}}
\newcommand{\CC}{\mathcal{C}}
\newcommand{\SSS}{\mathcal{S}}

\begin{document}

\title{Unbinned extraction of $\gamma$ from $B\to DK$ with normalizing flows}

\author[a]{Yuval~Grossman,}
\author[b,c]{Tony Menzo,}
\author[d]{Stefan Schacht,}
\author[a]{Chinhsan Sieng,}
\author[e]{and Jure Zupan}

\affiliation[a]{LEPP, Department of Physics, Cornell University, Ithaca, NY 14853, USA}
\affiliation[b]{Department of Physics and Astronomy, University of Alabama, Tuscaloosa, AL 35487, USA}
\affiliation[c]{Theoretical Physics Department, Fermilab, Batavia, Illinois 60510, USA}
\affiliation[d]{Institute for Particle Physics Phenomenology, Department of Physics, Durham University, Durham DH1 3LE, United Kingdom}
\affiliation[e]{Department of Physics, University of Cincinnati, Cincinnati, Ohio 45221, USA}

\emailAdd{yg73@cornell.edu}
\emailAdd{amenzo@ua.edu}
\emailAdd{stefan.schacht@durham.ac.uk}
\emailAdd{cs2284@cornell.edu}
\emailAdd{zupanje@ucmail.uc.edu}

\date{\today}

\preprint{IPPP/26/37, FERMILAB-PUB-26-0322-T}

\abstract{
We introduce an unbinned method for extracting the CKM angle $\gamma$ from the decay chain $B^\pm \to (D \to K_S \pi^+ \pi^-) K^\pm$ using normalizing flows (NFs). The NFs, trained on $D$ decay data, learn a faithful continuous representation of the amplitude and strong phase variation over the $D\to K_S\pi^+\pi^-$ Dalitz plot whose fidelity improves with increased data sample sizes. With this input, the $B$ decay data can be used to extract the parameters $r_B$, $\delta_B$, and $\gamma$.
We test the method on Monte Carlo generated data, where it successfully recovers the injected value of $\gamma$ within uncertainties. The present implementation propagates statistical uncertainties from finite training data via an ensemble of independently trained flows, and does not attempt to capture the effects of systematic experimental errors. We explore two versions of the method that differ in how the trigonometric constraint on phase variation is encoded, and comment on the possible extension to Bayesian NFs, which would provide direct uncertainty estimates on the learned densities without requiring ensemble training. 
}

\maketitle

\section{Introduction}
\label{sec:intro}

The Cabibbo--Kobayashi--Maskawa (CKM) unitarity triangle angle $\gamma$ (also denoted $\phi_3$, see, e.g., Refs.~\cite{Zupan:2019uoi,Grossman:2017thq}), 
\begin{equation}
\label{eq:gdef}
    \gamma = \arg\pqty{-\frac{V_{ud} V_{ub}^*}{V_{cd} V_{cb}^*}},
\end{equation}
can be extracted cleanly from $B^\pm \to D K^\pm$ decays through the interference of the tree-level transitions $b \to c\bar us$ and $b \to u\bar cs$ and their conjugates. This strategy has been understood for decades~\cite{Carter:1980hr,Carter:1980tk,Gronau:1990ra,Gronau:1991dp}. Its theoretical uncertainty is exceedingly small~\cite{Brod:2013sga,Brod:2014qwa}, so that in practice, the precision on the extracted value of $\gamma$ is limited by the amount of available experimental data, and on how this is used in the analysis.

The most precise current determinations of $\gamma$ rely on the BPGGSZ method~\cite{Bondar:2002,Giri:2003ty,Belle:2004bbr,Bondar:2005ki,Ceccucci:2020cim}, in which $D^0$ and $\widebar{D}^0$ decay to self-conjugate multibody final states such as $K_S\pi^+\pi^-$. In the model-independent implementation, the Dalitz plane is partitioned into bins and the charm inputs are encoded in bin-averaged observables. Optimized binning schemes~\cite{Bondar:2007ir,Bondar:2008hh} preserve most of the available sensitivity, but introduce a controlled coarse graining of the phase space. Given that the determination of $\gamma$ is mainly limited by statistical sizes of 
$B^\pm \to DK^\pm$ samples at LHCb and Belle~II~\cite{LHCb:2021dcr,LHCb:2020yot,Belle:2004bbr}, it would be desirable to recover this loss of local phase-space information via an appropriate unbinned version of the BPGGSZ method.

Several unbinned alternatives have been proposed. Ref.~\cite{Poluektov:2017zxp} replaced the choice of binning with a Fourier expansion in the strong phase difference, while Ref.~\cite{Lane:2023iak} proposed  instead an expansion in Legendre polynomials. Ref.~\cite{Backus:2022xhi}  formulated the problem in terms of cumulative distributions and a nonparametric test statistic. While the above approaches differ in the implementations, they do share a common goal of retaining more of the Dalitz-plot information than is the case for the widely adopted model-independent method based on finite binning~\cite{Bondar:2002,Giri:2003ty,Belle:2004bbr,Bondar:2005ki,Ceccucci:2020cim,Bondar:2007ir,Bondar:2008hh}. In the latter approach some loss of sensitivity to $\gamma$ occurs due to the averaging of the strong phase difference over data that is in a particular bin.

We pursue a complementary strategy based on density estimation with Normalizing Flows (NFs)~\cite{dinh2015, Kobyzev_2021, Rezende:2015a} (for a brief overview of NFs, see Appendix~\ref{app:nfs}). Already in the original BPGGSZ papers \cite{Giri:2003ty,Belle:2004bbr} it was pointed out that, if the variation of the $D$ decay amplitude $A_D e^{i\delta_D}$ across the Dalitz plot were known, it could be straightforwardly used to extract $\gamma$ from measurements of $B^\pm\to D(\to K_S\pi^+\pi^-) K^\pm$ decays.   At the time, implementing this idea required modeling of the $D$ decay amplitude using isobar ans\"atze, which introduced  irreducible modeling errors. As a result,  the binned model-independent method~\cite{Giri:2003ty} became a preferred experimental approach, despite some loss of sensitivity to $\gamma$. 

In contrast, the substantial progress in Machine Learning (ML) techniques for density estimation over the past two decades now makes it possible to revisit the original unbinned strategy.  As we will show, the NF based description of the $D$ decay amplitude leads to unbiased $\gamma$ determination. Moreover, unlike the available models for the Dalitz plot, such as the isobar model, the accuracy of the NF approach systematically improves as additional $D$ decay data becomes available, in close analogy with the increased precision of the $c_i$ and $s_i$ parameters in the model independent method when larger $D$ decay samples are used.

The application of NFs to $\gamma$ extractions does, however, face a technical challenge: for numerical stability, it is advantageous to learn from data the continuous densities of $A_D$, $A_D\cos \delta_D$ and $A_D\sin \delta_D$, rather than those of $A_D$ and $\delta_D$ directly. The reason is two-fold: {\em i)} in regions where $A_D$ is small, experimental sensitivity to the strong phase $\delta_D$ is poor, and {\em ii)} the observables entering the $\gamma$ extraction depend precisely on the combinations $A_D$, $A_D\cos\delta_D$, and $A_D\sin\delta_D$. The complication is that these three functions are not independent but rather obey a trigonometric identity. In this manuscript, we show how such a constraint can be consistently incorporated into the ML architecture.

The paper is structured as follows.
In \cref{sec:theory}, we review the dependence of the $B^\pm \to (K_S\pi^+\pi^-)_D K^\pm$ decay rates on $\gamma$, establish notation, summarize the binned BPGGSZ method and its limitations. In \cref{sec:unbinned_fit} we formulate two unbinned methods based on normalizing-flow density estimations. 
The simulation setup, including the amplitude model, datasets, and NF architectures, are described in \cref{sec:data}, while numerical results are shown 
 in \cref{sec:numerics}.
\Cref{sec:Implementations} contains a discussion of a path forward toward implementing the method with real data, followed by conclusions in \cref{sec:conclusions}. Appendix~\ref{app:nfs} contains further details about NF architecture and training, Appendix~\ref{app:constraint_aware} further details specifically about the $h-$network architecture, while Appendix~\ref{app:cp_coverage} contains additional results obtained using the 3-flow architecture.

\section{Determination of $\gamma$ from multibody decays}
\label{sec:theory}

We start by reviewing the common features of the methods that use $B^\pm\to D K^\pm$ decays to extract $\gamma$~\cite{Carter:1980hr,Carter:1980tk,Gronau:1990ra,Gronau:1991dp}, introducing the notation along the way, and reviewing the binned model independent BPGGSZ method.

\subsection{Notation and decay kinematics}
\label{sec:notation}
We focus specifically on 
\begin{equation*}
  B^\pm \to DK^{\pm} \to  ( K_S \pi^+ \pi^-)_DK^{\pm},
\end{equation*}
decay chain. The amplitudes for the decays receive two interfering contributions, from $B^+ \to D^0 K^+$ and $B^+ \to \widebar{D}^0 K^+$ transitions, giving 
\begin{align}
  \label{eq:amplitude}
{\cal A}\big(B^+ \!\to\! ( K_S \pi^+ \pi^-)_DK^{+}\big)&\!=\! A_B\Big[A_D(s_{12}, s_{13}) e^{i\delta (s_{12}, s_{13})} \!+\!r_B e^{i(\delta_B + \gamma)} A_D(s_{13}, s_{12}) e^{i\delta (s_{13}, s_{12})}\Big], \nonumber\\ 
{\cal A}\big(B^- \!\to\! ( K_S \pi^+ \pi^-)_DK^{-}\big)&\!=\! A_B\Big[A_D(s_{13}, s_{12}) e^{i\delta (s_{13}, s_{12})} \!+\!r_B e^{i(\delta_B - \gamma)} A_D(s_{12}, s_{13}) e^{i\delta (s_{12}, s_{13})}\Big].  
\end{align}
Here $A_B$ denotes the magnitude of the reference $B \to DK$ amplitude, taken to be real and positive by convention, 
\beq
  \mathcal{A}(B^- \to D^0 K^-) = \mathcal{A}(B^+ \to \widebar{D}^0 K^+) \equiv A_B,
\eeq
while $r_B= 0.0987\pm0.0022$~\cite{HeavyFlavorAveragingGroupHFLAV:2024ctg,HFLAV:GammaSummer2025} is the ratio of the suppressed to favored amplitudes, 
$\delta_B=(128.4\pm2.8)^\circ$~\cite{HeavyFlavorAveragingGroupHFLAV:2024ctg,HFLAV:GammaSummer2025} is their relative strong phase, $\gamma=(66.4\pm2.8)^\circ$~\cite{HeavyFlavorAveragingGroupHFLAV:2024ctg,HFLAV:GammaSummer2025} 
the weak phase, 
and
\beq
\mathcal{A}(B^- \to \widebar{D}^0 K^-) \equiv A_B r_B e^{i(\delta_B-\gamma)}, \quad \mathcal{A}(B^+ \to D^0 K^+) \equiv A_B r_B e^{i(\delta_B + \gamma)}.
\eeq
In the numerical pseudo-data analysis below we use, for simplicity, the following benchmark values,
\begin{align}
r_B = 0.1\,, \quad  \delta_B = 130^\circ\,, \quad \gamma = 70 ^\circ,
\end{align}
which agree with world averages within one sigma.
In \cref{eq:amplitude} the amplitudes for the three-body decay of the intermediate $D$ meson, 
\begin{equation*}
  D^0 \to K_S(p_1) \pi^+(p_2) \pi^-(p_3),
\end{equation*}
are denoted as
\begin{equation}
  \mathcal{A}(D^0 \to K_S \pi^+ \pi^-) \equiv A_D(s_{12}, s_{13}) e^{i\delta (s_{12}, s_{13})}\;,
\end{equation}
\begin{equation}
  \mathcal{A}(\widebar{D}^0 \to K_S \pi^+ \pi^-) \equiv A_D(s_{13}, s_{12}) e^{i\delta (s_{13}, s_{12})}. 
\end{equation}
Note that the amplitude for $\mathcal{A}(\widebar{D}^0 \to K_S \pi^+ \pi^-)$ is obtained via a CP transformation, assuming no CP violation in $D$ decays. Here,  $p_1,p_2,p_3$ are the four-momenta of $K_S, \pi^+, \pi^-$, respectively, while $s_{ij} \equiv (p_i + p_j)^2$ is the invariant mass squared of the $ij$ system.  The magnitude of the decay amplitude, $A_D$, and the strong phase, $\delta$, are real functions  that vary across the Dalitz plot, satisfying $A_D \geq 0$ and $\delta \in [0,2\pi]$. 

In the numerical analysis, we use the square-Dalitz plot coordinates $(m',\theta')$, obtained via smooth bijective transformation of the physical Dalitz plot to the unit square~\cite{BaBar:2005jqu,BaBar:2007hmp,Back:2017zqt}
\begin{equation}
\label{eq:m':theta'}
  m' \equiv \frac{1}{\pi} \arccos \left( 2 \frac{\sqrt{s_{23}} - m_{\text{min}}}{m_{\rm max} - m_{\rm min}} - 1 \right), \qquad 
  \theta' \equiv \frac{\theta_{23}}{\pi} \;,
\end{equation}
where 
\beq
m_{\rm min} = 2 m_\pi, \qquad m_{\rm max} = m_D - m_K,
\eeq
 and $\theta_{23} \in [0,\pi]$ is the helicity angle between the $\pi^-$ and $K_S$ three-momenta in the di-pion rest frame, i.e.,
\begin{equation}
  \cos \theta_{23} = \hat{n}_{\pi^-} \cdot \hat{n}_{K_S}, \quad \text{where} \quad \hat{n}_i \equiv \frac{\vec{p}^{\,*}_i}{|\vec{p}^{\,*}_i|}\;,
\end{equation}
with $\vec{p}^{\,*}_i$ the three-momentum of particle $i=\pi^-, K_S$  in the di-pion rest frame.
The $m', \theta'$ variables are normalized in such a way that they both lie in a unit interval, $m',\theta' \in [0,1]$.

In terms of square Dalitz plot variables the two $D$ decay amplitudes are given by,
\begin{equation}
  \mathcal{A}(D^0 \to K_S \pi^+ \pi^-) \equiv A_D(m', \theta') e^{i\delta (m', \theta')}\;,
\end{equation}
\begin{equation}
  \mathcal{A}(\widebar{D}^0 \to K_S \pi^+ \pi^-) \equiv \bar A_D(m', \theta') e^{i\bar\delta (m', \theta')} = A_D(m', 1-\theta') e^{i\delta (m', 1-\theta')}\;.
\end{equation}
That is, the CP transformation which exchanges $\pi^+ \leftrightarrow \pi^-$ transforms  $m' \to m'$ ($s_{23} = (p_2 + p_3)^2$ is invariant under $p_2 \leftrightarrow p_3$) and $\theta' \to 1-\theta'$ ($\theta_{23} \to \pi - \theta_{23}$ under $p_2 \leftrightarrow p_3$), corresponding to the $s_{12} \leftrightarrow s_{13}$ exchange in the standard Dalitz coordinates. 

The reduced partial decay widths, i.e., the partial decay widths for  $B^\pm \to( K_S \pi^+ \pi^-)_DK^{\pm}$ decays with $A_B^2$ and phase space factors taken out, 
 are in the square Dalitz plot variables given by 
\begin{align}
\frac{d\hat{\Gamma}_{+}}{d\Phi_3} &= 
  A_D^2 + r_B^2\, \bar A_D^2 
  + 2r_B\, A_D\, \bar A_D \cos(\delta_B + \gamma - \Delta\delta),
  \label{eq:hat_Gamma_plus}
  \\
  \frac{d\hat{\Gamma}_{-}}{d\Phi_3} &= 
  \bar A_D^2 + r_B^2\, A_D^2 
  + 2r_B\, A_D\, \bar A_D \cos(\delta_B - \gamma + \Delta\delta),
  \label{eq:hat_Gamma_minus} 
\end{align}
where $d\Phi_3 =|\det J|\,dm'\,d\theta'$ is the infinitesimal phase space element expressed in square Dalitz plot variables. In Eqs.~\eqref{eq:hat_Gamma_plus}, \eqref{eq:hat_Gamma_minus}, 
we do not show explicitly the dependence on the square Dalitz plot variables, i.e., we shortened
$\bar A_D=\bar A_D(m', \theta')$, $A_D=A_D(m', \theta')$, and similarly for the strong-phase difference between the CP-conjugate Dalitz points,
\begin{equation}
    \Delta\delta(m',\theta') \equiv\delta(m',\theta') - \bar\delta(m',\theta').
\end{equation}
The Jacobian factor due to the change of variables is~\cite{Back:2017zqt}
\begin{equation}
  \left|\det J(m',\theta')\,\right| = \frac{\pi^2}{2}\,\Delta m 
  \sqrt{\lambda\!\left(m_D^2,\, m_{K_S}^2,\, s_{23}\right)} 
  \sqrt{s_{23} - 4 m_\pi^2} 
  \;\sin(\pi m')\sin(\pi \theta'),
\end{equation}
where $\Delta m = m_{\rm max} - m_{\rm min}$, and $\lambda(a,b,c) = 
a^2+b^2+c^2-2ab-2bc-2ca$ is the K\"all\'en function.

\subsection{Binned BPGGSZ method}
\label{sec:binned}
The weak phase $\gamma$ changes sign in the interference term for  $B^+ \to( K_S \pi^+ \pi^-)_DK^+$ and $B^- \to( K_S \pi^+ \pi^-)_DK^-$ decay widths, cf. Eqs. \eqref{eq:hat_Gamma_plus}  and \eqref{eq:hat_Gamma_minus}. By measuring these partial decay widths one can therefore determine $\gamma$ (together with $r_B$ and $\delta_B$) as long as $A_D$, $\bar A_D$ and $\Delta \delta$ are measured separately in $D$ decays. There are two approaches to this: {\em i)} the {\em model-dependent} method \cite{Giri:2003ty,Belle:2004bbr}, in which $A_D(m',\theta')$ and $\delta(m',\theta')$ are determined from a fit of $D$ decay data to an isobar model, and {\em ii)} the {\em model-independent} method \cite{Giri:2003ty}, in which the $D$ decay data is binned, giving the per-bin averaged values of the $D$ decay amplitude and strong phase. We review the model-independent method in the remainder of this subsection.

In the model-independent method, the Dalitz plane is partitioned into $2N$ CP conjugate bins, i.e., the bins are pair-wise related  by the $\pi^+\leftrightarrow\pi^-$ exchange which transforms bin $i$ to bin $\bar i$. Adapting the notation of Ref. \cite{Bondar:2008hh} we then define the integrated $D$ and $\bar D$ decay amplitudes squared for bin $i$, 
\begin{equation}
\label{eq:Ki:def}
K_i = \int_{\text{bin }i} A_D^2(m',\theta') \, d\Phi_3, \qquad 
\bar K_i = \int_{\text{bin }i} \bar A_D^2(m',\theta') \, d\Phi_3,
\end{equation}
as well as the weighted averages of the cosine and sine of the strong phase difference,\footnote{Note that the normalization of the $c_i$ and $s_i$ is different from the one in Ref.~\cite{Giri:2003ty}.}
\begin{align}
\label{eq:ci:def}
  c_i &= \frac{1}{\sqrt{K_i \bar K_i}} \int_{\text{bin }i} A_D(m',\theta')\,\bar A_D(m',\theta') \cos\Delta\delta(m',\theta') \, d\Phi_3,
\\
\label{eq:ci:def}
  s_i &= \frac{1}{\sqrt{K_i \bar K_i}} \int_{\text{bin }i} A_D(m',\theta')\,\bar A_D(m',\theta') \sin\Delta\delta(m',\theta') \, d\Phi_3.
\end{align}
Note that in the limit of negligible CP violation in $D$ decays, which is the limit we work in,
one has $K_{\bar i}=\bar K_i$, $c_i=c_{\bar i}$, $s_i=-s_{\bar i}$. 
 (The effects of nonzero CP violation in $D$ decays, as well as the effects of $D-\bar D$ mixing can be included~\cite{Grossman:2005rp,
 Martone:2012nj,Rama:2013voa}.)
The $K_i$, $\bar K_i$
can be measured with high precision using the available large samples of flavor-tagged $D \to K_S \pi^+ \pi^-$ decays~\cite{CLEO:2002uvu,Belle:2007tti,BaBar:2010nhz,Belle:2014ydf,Reichert:2013ewa}, while $c_i$ and $|s_i|$ 
can be determined from quantum-correlated $\psi(3770)\to D\bar D$ decays at CLEO-c~\cite{CLEO:2010iul} and BESIII~\cite{BESIII:2020khq}.

Belle reconstructed $\approx 1.2 \times 10^6$ $D^*$-tagged $D^0 \to K_S \pi^- \pi^+$ decays at an integrated luminosity of 921 fb$^{-1}$ with sample purity at 96\% \cite{Belle:2014ydf}. Event count projections for a similar analysis at Belle II can be conservatively estimated to be $\approx 6.7 \times 10^6$ at 5 ab$^{-1}$, $\approx 26.7 \times 10^6$ at 20 ab$^{-1}$, etc. 
Information on the strong phase was first extracted by CLEO-c using 0.82 fb$^{-1}$ of data at the $\psi(3770)$ resonance by analyzing $D\to K_S\pi^-\pi^+$ directly, corresponding to $\mathcal{O}(10^3)$ usable double-tag events distributed among CP-tags and double Dalitz modes \cite{CLEO:2010iul}. 
BESIII subsequently analyzed 2.93 fb$^{-1}$ at $\psi(3770)$, reconstructing several thousand CP-tagged $K_S\pi^-\pi^+$ events (augmented by partial-reconstruction of $K_L\pi^-\pi^+$ tags) and on the order of a few hundred fully reconstructed double-Dalitz events \cite{BESIII:2020khq}. 
A more recent BESIII update based on 7.93 fb$^{-1}$ reports approximately $10^4$ CP-tagged 
$K_S\pi^-\pi^+$ events and $\sim 5\times 10^3$ fully reconstructed $(K_S\pi^-\pi^+)^2$ events, 
together with a comparable number of $K_S\pi^-\pi^+$ vs.\ $K_L\pi^-\pi^+$ events~\cite{BESIII:2025nsp}.
With the complete BESIII dataset of about $20$ fb$^{-1}$ accumulated at the charm threshold~\cite{Gilman:2026nys}, naive luminosity rescaling predicts $\mathcal{O}(\text{a few} \times 10^4)$ CP-tagged $K_S\pi^-\pi^+$ events and $\mathcal{O}(10^4)$ reconstructed double-Dalitz events, representing roughly a factor of 20 or more increase over CLEO-c statistics. 

Here, the observed event counts are proportional to~\cite{Bondar:2008hh, BaBar:2010nhz}
\begin{equation}
\label{eq:countformula}
  N_{i} \;\propto\; K_i + \bar K_i \pm 2 \sqrt{K_i \bar K_i}\,c_i,
\end{equation}
for CP-even (CP-odd) tagged decays, and to
\begin{equation}
  N_{ij} \;\propto\; K_i \bar K_j + \bar K_i K_j - 2 \sqrt{K_i \bar K_i K_j \bar K_j}\,(c_i c_j + s_i s_j),
\end{equation}
for double-Dalitz decays, i.e., for the case where both $D$ and $\bar D$ in $\psi(3770)\to D\bar D$  decay via $D\to K_S\pi^-\pi^+$. The overall sign ambiguity in the determination of $s_i$ can be resolved by using the  information from the isobar model, which for this purpose is more than precise enough.

The same bin choice is then used also in $B^\pm \to( K_S \pi^+ \pi^-)_DK^{\pm}$ decays, with partial decay widths in bin $i$ given by, cf. Eqs.~\eqref{eq:hat_Gamma_plus}, \eqref{eq:hat_Gamma_minus}.  
\begin{align}
\hat{\Gamma}_{B^+,i} &= 
  K_i + r_B^2\, \bar K_i
  + 2r_B\, \sqrt{K_i \bar K_i} \big[ \cos(\delta_B + \gamma)c_i + \sin(\delta_B + \gamma)s_i\big],
  \label{eq:hat_Gamma_plus:binned}
  \\
  \hat{\Gamma}_{B^-,i} &= 
 \bar K_i + r_B^2\, K_i
  + 2r_B\, \sqrt{K_i \bar K_i} \big[ \cos(\delta_B - \gamma)c_i - \sin(\delta_B - \gamma)s_i\big],
  \label{eq:hat_Gamma_minus:binned} 
\end{align}
Since $K_i$, $\bar K_i$, $c_i$, $s_i$ are known from $D$ decays, one can now extract $\gamma$ from $B$ decays. However, if the strong phase difference varies rapidly over the bin, this would lead to   $c_i, s_i \ll 1$, and reduced sensitivity to $\gamma$. To minimize this effect, experimental analyses use 
$N=8$ bin pairs 
with optimized binning schemes following contours of $\Delta\delta(m',\theta')$ 
predicted by the isobar model~\cite{Bondar:2008hh,BESIII:2020khq}. This improves statistical efficiency considerably, but does not eliminate the washout due to underlying coarse graining entirely. 
This trade-off motivates the unbinned approaches.

\section{Unbinned method via normalizing flows}
\label{sec:unbinned_fit}

The unbinned method based on NFs 
follows the same logic as the model independent binned BPGGSZ analysis, but replaces the bin-integrated inputs from $D$ decays, $K_i,\bar K_i,c_i,s_i$ with the corresponding continuous functions over the square Dalitz plot. That is, in analogy with Eqs. \eqref{eq:Ki:def}-\eqref{eq:ci:def} we define 
\begin{align}
\label{eq:K_def}
    K(m',\theta') &\equiv \frac{|\det J| }{\hat \Gamma_{K_S\pi\pi}} \, |A_D(m',\theta')|^2,
\\
\label{eq:C_def}
    \mathcal{C}(m',\theta') & \equiv  \sqrt{K(m',\theta') \bar{K}(m',\theta')} \cos\Delta\delta(m',\theta'),
\\
\label{eq:S_def}
    \mathcal{S}(m',\theta') & \equiv \sqrt{K(m',\theta') \bar{K}(m',\theta')} \sin\Delta\delta(m',\theta'),
\end{align}
where
\beq
\hat \Gamma_{K_S\pi\pi}= \int |A_D|^2 \, d\Phi_3= \int |A_D|^2 \, |\det J| dm'd\theta',
\eeq
 is the reduced partial $D \to K_S\pi^+\pi^-$ decay width. The prefactor $|\det J|/\hat \Gamma_{K_s \pi\pi}$ in the definition  of  $K(m',\theta')$ ensures that this is a probability density in the square Dalitz plot, and can thus be learned from data using density estimators such as NFs. We use normalizing flows because they provide flexible density models with exact likelihood evaluation (a short summary is given in Appendix~\ref{app:nfs}).
Even though $\mathcal{C}(m',\theta')$ and  $\mathcal{S}(m',\theta')$ are not probability distributions they can still be learned from data using machine learning, as we explain below. 

 The reduced partial $B^\pm \to (K_S\pi^+\pi^-)_D K^\pm$ decay widths are given by, cf.
Eqs.~\eqref{eq:hat_Gamma_plus}, \eqref{eq:hat_Gamma_minus},
\begin{align}
  \frac{d\hat{\Gamma}_{B^+}}{dm'\,d\theta'} &= \Gamma_{K_S\pi\pi}  \left\{ K + r_B^2 \bar{K} + 2 r_B \left[ \cos(\delta_B + \gamma) \, \mathcal{C} + \sin(\delta_B + \gamma) \, \mathcal{S} \right] \right\}, 
  \label{eq:dgamma_plus_obs}
  \\
    \frac{d\hat{\Gamma}_{B^-}}{dm'\,d\theta'} &= \Gamma_{K_S\pi\pi} \left\{ \bar{K} + r_B^2 K + 2 r_B \left[ \cos(\delta_B - \gamma) \, \mathcal{C} - \sin(\delta_B - \gamma) \, \mathcal{S} \right] \right\}, 
    \label{eq:dgamma_minus_obs}
\end{align}
where for notational similarity with the unbinned method we denoted the density at the CP-conjugate point as
\begin{equation}
\label{eq:barK:def}
    \bar{K}(m',\theta') \equiv K(m',1-\theta').
\end{equation}

Note that the functions in Eqs. \eqref{eq:K_def}--\eqref{eq:S_def} satisfy the constraint
\begin{equation}\label{eq:CS_constraint}
    \mathcal{C}^2(m',\theta') + \mathcal{S}^2(m',\theta') = K(m',\theta') \bar{K}(m',\theta'),
\end{equation}
which is nothing more than the trigonometric relation $\cos^2(x)+\sin^2(x)=1$, but written in our notation. That is, there are only two independent functions, 
$ K$ and $\mathcal{C}$
that need to be learned from $D$ decay data, while  
$\mathcal{S}$
is given by 
\begin{equation}
\label{eq:CS_constraint}
    \mathcal{S}(m',\theta') = \pm \sqrt{K(m',\theta') \bar{K}(m',\theta')- \mathcal{C}^2(m',\theta')},
\end{equation}
where the sign cannot be determined 
from data alone.
 The fit to the isobar model can be used to resolve this ambiguity. 

\begin{figure*}[t!]
\centering
\resizebox{0.96\textwidth}{!}{%
\begin{tikzpicture}[
    >=Stealth,
    font=\footnotesize,
    node distance=0.7cm and 1.3cm,
    data/.style={
        draw=MidnightBlue!50,
        rounded corners=5pt,
        fill=MidnightBlue!6,
        minimum height=0.90cm,
        minimum width=2.55cm,
        inner sep=4pt,
        align=center
    },
    flow/.style={
        draw=BurntOrange!60!black,
        rounded corners=5pt,
        fill=BurntOrange!10,
        minimum height=0.90cm,
        minimum width=2.2cm,
        inner sep=4pt,
        align=center
    },
    obs/.style={
        draw=ForestGreen!50!black,
        rounded corners=5pt,
        fill=ForestGreen!8,
        minimum height=0.90cm,
        minimum width=2.1cm,
        inner sep=4pt,
        align=center
    },
    netbox/.style={
        draw=Violet!60!black,
        rounded corners=5pt,
        fill=Violet!8,
        minimum height=0.90cm,
        minimum width=2.2cm,
        inner sep=4pt,
        align=center
    },
    fitbox/.style={
        draw=BrickRed!55!black,
        rounded corners=5pt,
        fill=BrickRed!7,
        minimum height=0.95cm,
        minimum width=2.35cm,
        inner sep=4pt,
        align=center
    },
    result/.style={
        draw=Goldenrod!60!black,
        very thick,
        rounded corners=5pt,
        fill=Goldenrod!10,
        minimum height=0.95cm,
        minimum width=1.90cm,
        inner sep=4pt,
        font=\footnotesize\bfseries,
        align=center
    },
    arr/.style={->, very thick, draw=black!60, rounded corners=5pt},
    stagebox/.style={
        draw=black!15,
        rounded corners=8pt,
        fill=black!2,
        inner sep=7pt
    },
    stagelabel/.style={
        font=\scriptsize\bfseries\sffamily,
        text=black!50,
        fill=white,
        inner sep=2pt
    },
    panellabel/.style={
        font=\normalsize\bfseries,
        text=black!80
    },
]


\def\voff{-7.6cm}

\node[data] (flav2) {Flavor-tagged\\[-1pt]$D\to K_S\pi^+\pi^-$};
\node[flow, right=1.3cm of flav2] (fflow2) {Flavor\\[-1pt]NF};
\node[obs, right=1.3cm of fflow2] (K2) {$K(m',\theta')$};

\node[data, below=0.90cm of flav2] (cptag2) {CP-tagged\\[-1pt]$\psi(3770)\to D\bar D$};
\node[netbox] (hnet) at (cptag2 -| fflow2) {MLP};
\node[obs] (hobs) at (cptag2 -| K2) {$h(m',\theta')$};

\coordinate (fitmid2) at ($(K2)!0.5!(hobs)$);
\node[fitbox] (like2) at ($(fitmid2)+(4.1,-0.10)$) {Unbinned\\[-1pt]likelihood fit};
\node[data, above=0.60cm of like2] (bdata2) {$B^\pm\to DK^\pm$\\[-1pt]data};
\node[result, right=0.85cm of like2] (gamma2) {$(r_B,\delta_B,\gamma)$};

\draw[arr] (flav2) -- (fflow2);
\draw[arr] (fflow2) -- (K2);
\draw[arr] (cptag2) -- (hnet);
\draw[arr] (hnet) -- (hobs);

\draw[arr] (K2.south) -- (hnet.north);

\draw[arr] (K2.east) -- ++(0.55,0) |- ([yshift=0.16cm]like2.west);
\draw[arr] (hobs.east) -- ++(0.55,0) |- ([yshift=-0.16cm]like2.west);

\draw[arr] (bdata2) -- (like2);
\draw[arr] (like2) -- (gamma2);

\begin{scope}[on background layer]
    \node[stagebox, fit=(flav2)(cptag2)] (s1b) {};
    \node[stagebox, fit=(fflow2)(hnet)] (s2b) {};
    \node[stagebox, fit=(K2)(hobs)] (s3b) {};
    \node[stagebox, fit=(bdata2)(like2)(gamma2)] (s4b) {};
\end{scope}

\node[stagelabel] at (s1b.north) {Charm inputs};
\node[stagelabel] at (s2b.north) {Density estimation};
\node[stagelabel] at (s3b.north) {Learned quantities};
\node[stagelabel] at (s4b.north) {$B$-decay fit};

\node[panellabel, anchor=south west] at ($(s1b.north west)+(-0.1,0.25)$) {(a) Constraint-aware architecture};


\node[data] (flav) at ($(flav2)+(0,70\voff)$) {Flavor-tagged\\[-1pt]$D\to K_S\pi^+\pi^-$};
\node[data, below=0.65cm of flav] (cpeven) {CP-even tagged\\[-1pt]$\psi(3770)\to D\bar D$};
\node[data, below=0.65cm of cpeven] (cpodd) {CP-odd tagged\\[-1pt]$\psi(3770)\to D\bar D$};

\node[flow, right=1.3cm of flav] (fflow) {Flavor\\[-1pt]NF};
\node[flow, right=1.3cm of cpeven] (eflow) {CP-even\\[-1pt]NF};
\node[flow, right=1.3cm of cpodd] (oflow) {CP-odd\\[-1pt]NF};

\node[obs, right=1.3cm of fflow] (K) {$K(m',\theta')$};
\coordinate (cpmid) at ($(eflow)!0.5!(oflow)$);
\node[obs] (C) at (K |- cpmid) {$\mathcal{C}(m',\theta')$};
\node[obs] (S) at ($(K)!0.5!(C)$) {$\mathcal{S}(m',\theta')$};

\coordinate (fitmid) at (S);
\node[fitbox] (like) at ($(fitmid)+(3.7,-0.10)$) {Unbinned\\[-1pt]likelihood fit};
\node[data, above=0.60cm of like] (bdata) {$B^\pm\to DK^\pm$\\[-1pt]data};
\node[result, right=0.85cm of like] (gamma) {$(r_B,\delta_B,\gamma)$};

\draw[arr] (flav) -- (fflow);
\draw[arr] (cpeven) -- (eflow);
\draw[arr] (cpodd) -- (oflow);

\draw[arr] (fflow) -- (K);
\draw[arr] (eflow.east) -- ++(0.35,0) |- (C.west);
\draw[arr] (oflow.east) -- ++(0.35,0) |- (C.west);

\draw[arr] (K) -- (S);
\draw[arr] (C) -- (S);

\draw[arr] (K.east) -- ++(0.35,0) |- ([yshift= 0.20cm]like.west);
\draw[arr] (S.east) -- (like.west);
\draw[arr] (C.east) -- ++(0.35,0) |- ([yshift=-0.20cm]like.west);

\draw[arr] (bdata) -- (like);
\draw[arr] (like) -- (gamma);

\begin{scope}[on background layer]
    \node[stagebox, fit=(flav)(cpodd)] (s1a) {};
    \node[stagebox, fit=(fflow)(oflow)] (s2a) {};
    \node[stagebox, fit=(K)(S)(C)] (s3a) {};
    \node[stagebox, fit=(bdata)(like)(gamma)] (s4a) {};
\end{scope}

\node[stagelabel] at (s1a.north) {Charm inputs};
\node[stagelabel] at (s2a.north) {Density estimation};
\node[stagelabel] at (s3a.north) {Learned quantities};
\node[stagelabel] at (s4a.north) {$B$-decay fit};

\node[panellabel, anchor=south west] at ($(s1a.north west)+(-0.1,0.25)$) {(b) Deferred constraint enforcement};

\end{tikzpicture}%
}
\caption{Schematic of the two analysis pipelines. \textbf{(a)} {\em Constraint-aware architecture}: the flavor NF  is trained first to learn $K(m',\theta')$, which then serves also as an input for the network $h:[0,1]^2\to[-1,1]$ trained on CP-tagged data; the pair $(K,h)$ determines $\mathcal{C}=\sqrt{K\bar{K}}\,h$ and $|\mathcal{S}|=\sqrt{K\bar{K}}\sqrt{1-h^2}$, satisfying the trigonometric constraint, \cref{eq:constraint:C}, by construction.
\textbf{(b)} {\em Deferred constraint enforcement}: three independent NFs learn $K(m',\theta')$, $p_+(m',\theta')$, and $p_-(m',\theta')$; the interference observable $\mathcal{C}$ is extracted algebraically;
$\mathcal{S}$ is obtained from $\mathcal{C}^2+\mathcal{S}^2=K\bar{K}$ with the sign determined from the amplitude model.
Both pipelines use the same unbinned maximum-likelihood fit for $(r_B,\delta_B,\gamma)$ from $B$ decay data. 
}
\label{fig:pipeline_combined}
\end{figure*}
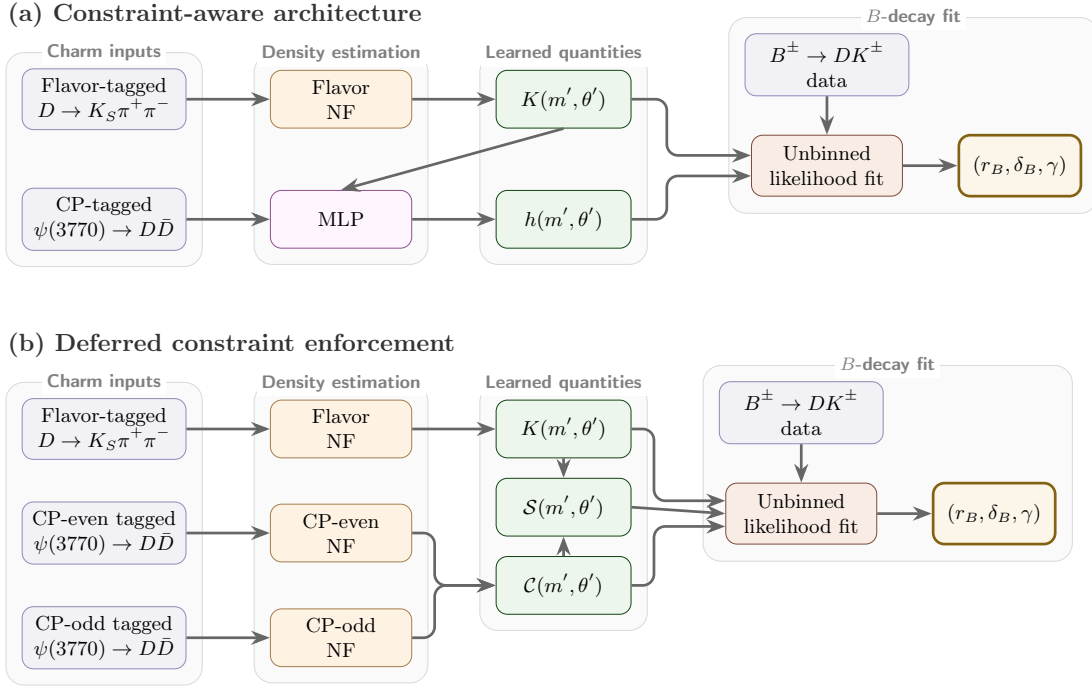

Note that the function $\mathcal{C}(m',\theta')$ is not completely free, as it needs to satisfy the constraint
\beq
\label{eq:constraint:C}
\big|\mathcal{C}(m',\theta')\big|\leq \sqrt{ K(m',\theta') \bar{K}(m',\theta') },
\eeq
while $K(m',\theta')$ is required only to be non-negative and to integrate to 1 over the square Dalitz plot. The constraint \eqref{eq:constraint:C} makes the problem of learning $K(m',\theta') $ and $\mathcal{C}(m',\theta')$ from $D$ decay data nontrivial. We pursue two strategies, as described below and illustrated schematically in Fig.~\ref{fig:pipeline_combined}.

\subsection{Constraint-aware architecture (\texorpdfstring{$h-$}{h-}network)} A normalizing-flow representation of $K(m',\theta')$ is first trained on flavor-tagged
Monte-Carlo samples for $D \to K_S \pi^+\pi^-$ decays.
For $K_\text{NF}$ we use a rational-quadratic neural spline flow (RQ-NSF)~\cite{Durkan:2019nsq}, as implemented in the \textsc{nflows} library~\cite{nflows}; the architecture and training are described in Appendix~\ref{app:nfs}.
The obtained $K_\text{NF}(m',\theta')$ is then frozen and used in the second step, where we define an ML surrogate 
\beq
\label{eq:h:definition}
\mathcal{C}(m',\theta')= \sqrt{ K(m',\theta') \bar{K}(m',\theta') }\times h(m',\theta'),
\eeq
where $h:[0,1]^2\to[-1,1]$ is a regression network with a $\tanh$ output activation, so that $|h|\leq 1$ by construction. That is, $h$ is a neural network representation of $\cos\Dd$, cf. \cref{eq:C_def}.
Because $\cos\Dd$ varies rapidly near narrow resonances, we use a Sinusoidal Representation Network (SIREN)~\cite{Sitzmann2020siren} with sinusoidal hidden-layer activations, which naturally captures oscillatory structures across the Dalitz plot. 
With the square-Dalitz parametrization in which $m'$ is constructed from the $\pi^+\pi^-$ invariant mass, $\cos\Dd$ is symmetric under $\theta'\to 1-\theta'$. 
We enforce this symmetry by averaging the raw network output over $(m',\theta')$ and $(m',1-\theta')$. The full SIREN architecture, the loss function, and training details are given in Appendix~\ref{app:constraint_aware}.

The network $h$ is trained on CP-tagged $\psi(3770)\to D_\text{CP} (D\to K_S\pi^+\pi^-)$ Monte-Carlo generated decay data,  where for CP-even (CP-odd) tag final states, the distribution of $D\to K_S\pi^+\pi^-$ decays is given by, see Eq.~(\ref{eq:countformula}): 
\beq
\label{eq:Gammapm:h}
    \frac{d\Gamma_\pm}{dm' d\theta'} \propto K + \KK \pm 2\sqrt{K\KK}\, h.
    \eeq
The frozen $K_\text{NF}$ enters this expression, and $h$ is optimized by minimizing the negative log-likelihood on the CP-tagged samples, with a normalization integral 
calculated via Monte Carlo (see Appendix~\ref{app:constraint_aware}). The resulting estimate for $\mathcal{C}$ is obtained from the r.h.s.\ of \cref{eq:h:definition}, which we denote as $\mathcal{C}_\text{NF}$ for convenience even though both the NF and SIREN are used in its computation.
In the actual experimental analysis a more complete set of entangled $\psi(3770)\to D \bar D$ decays can also be incorporated, including  $\psi(3770)\to   (D\to K_S\pi^+\pi^-) (\bar D\to K_S\pi^+\pi^-)$.

\subsection{Deferred constraint enforcement ($3-$flow)} An alternative approach is to train three independent normalizing flows without imposing any constraints. As in the constraint-aware architecture a {\em flavor NF} $K(m',\theta')$ is trained on flavor-tagged $D \to K_S \pi^+ \pi^-$ decay data, while additional {\em CP-even NF} and {\em CP-odd NF}  are trained on the corresponding CP tagged data. 
All three flows share the same RQ-NSF architecture~\cite{Durkan:2019nsq}, with details given in Appendix~\ref{app:nfs}.
That is, the two additional NFs are the density estimators for the probability distributions (see also App. \ref{sec:loss:function:norm})
\beq
\label{eq:p+-:def}
p_\pm(m',\theta')=\frac{1}{\hat{\Gamma}_\pm} \frac{d\hat{\Gamma}_\pm}{dm' d\theta'},
\eeq
allowing for an NF-based approximation of $\mathcal{C}$, denoted as $\mathcal{C}_\text{NF}(m',\theta')$, via the relation, see Eq.~(\ref{eq:countformula}): 
\begin{equation}
\label{eq:C_extraction}
    \mathcal{C}(m',\theta') = \frac{\hat{\Gamma}_+ \, p_+(m',\theta') - \hat{\Gamma}_- \, p_-(m',\theta')}{\hat{\Gamma}_+ + \hat{\Gamma}_-}.
\end{equation}

Since $K_\text{NF}$ and $\mathcal{C}_\text{NF}$ are now extracted from independently trained flows, statistical fluctuations can produce unphysical points with $\mathcal{C}_\text{NF}^2 > K_\text{NF}\bar{K}_\text{NF}$, violating \cref{eq:constraint:C}. In the final step the estimates for $\mathcal{C}$ and $K$ are then improved by averaging over $k$
nearest neighbors in the square Dalitz plot, increasing $k$ until the constraint \cref{eq:CS_constraint} is satisfied.

In both strategies, once NFs $K_\text{NF}(m',\theta')$ and ${\mathcal C}_\text{NF}(m',\theta')$ are obtained from $D$ decay data, they can subsequently be used in the interpretation of $B$ decay data, Eqs. \eqref{eq:dgamma_plus_obs}, \eqref{eq:dgamma_minus_obs}, together with Eqs. \eqref{eq:barK:def}, \eqref{eq:CS_constraint} to determine $\gamma$. Below, we demonstrate how this method performs numerically on simulated data for both strategies, the constraint-aware architecture and the deferred constraint enforcement.

\section{Simulation setup}
\label{sec:data}
Both the constraint-aware and deferred constraint architectures use the same simulated training datasets. Subsections~\ref{sec:amplitude:model} and~\ref{sec:simulated:data} describe the amplitude model and the simulated datasets, while subsection~\ref{sec:B:likelihood} introduces the unbinned likelihood for $\gamma$, $\delta_B$, and $r_B$. Architecture and training details for all networks are deferred to Appendices~\ref{app:nfs} and~\ref{app:constraint_aware}.

\subsection{Amplitude model for $D\to K_S\pi^+\pi^-$ decays}
\label{sec:amplitude:model}

All simulated samples of $D\to K_S\pi^+\pi^-$ decay data were generated using the isobar amplitude model for $D^0 \to K_S \pi^+ \pi^-$ described in Refs.~\cite{BaBar:2018cka,Back:2017zqt}, with the default values of the parameters. The model includes the $\pi\pi$ resonances $\rho(770)$, $\omega(782)$, $f_2(1270)$, $\rho(1450)$, the $K\pi$ resonances, $K^*(892)$, $K^*_2(1430)$, $K^*(1680)$, $K^*(1410)$, with Blatt--Weisskopf barrier factors, as well as a five-pole $K$-matrix for the $\pi\pi$ S-wave, and the LASS 
parameterization for the $K\pi$ S-wave.
(The LASS parametrization is the model of the elastic $K\pi$ S-wave based on LASS scattering data \cite{Aston:1987ir}.)
The isobar amplitude model was defined in terms of standard Dalitz variables $(s_{12}, s_{13})$ in Refs.~\cite{BaBar:2018cka,Back:2017zqt}. We mapped these definition to the square-Dalitz coordinates $(m', \theta')$, which were used in all the training and fitting steps of the analysis.

\subsection{Simulated datasets}
\label{sec:simulated:data}

The simulated datasets consist of three samples: a flavor-tagged $D \to K_S\pi^+\pi^-$ decay data sample, a CP-tagged $D \to K_S\pi^+\pi^-$ data sample, and the $B^\pm \to (D\to K_S\pi^+\pi^-) K^\pm$ decay data sample. All datasets are generated using the isobar amplitude model in Refs.~\cite{BaBar:2018cka,Back:2017zqt}. Below we describe each of the above simulated datasets in turn.

\paragraph{Flavor-tagged $D \to K_S\pi^+\pi^-$ sample.} The charge of the soft pion in $D^{*\pm}\to D^0\pi^\pm$ decays tags the flavor of the $D^0$ meson.  Such flavor-tagged $D^0 \to K_S\pi^+\pi^-$ events can  be used to learn the square Dalitz density, $K(m',\theta')$, defined in \cref{eq:K_def}. For the numerical analysis we generated $N_\text{ens}^D=25$ independent datasets, each containing an ensemble of $N_{\rm flav} = 2\times10^6$ events that follow the density distribution predicted by the isobar model (the datasets were generated with the accept-reject algorithm using sampling from a
uniform distribution over the physical Dalitz region). 
The events in each of the $N_\text{ens}^D$ datasets were further split into the randomly and uniformly selected training (90\% of events) and validation (10\%) subsets. 

\paragraph{CP-tagged samples.} The $D$ and $\bar D$ decays in the $\psi(3770)\to D\bar{D}$ decay chain are quantum entangled. If the $D$ decay on the tag side decays to a CP-even(CP-odd) final state, such as $D\to \pi^+\pi^-$ ($D\to K_S\pi^0$), this means that the decay on the signal side proceeds from the $\big( |D^0\rangle\pm |\bar D^0 \rangle \big)/\sqrt{2}$ wave function \cite{Gronau:2001nr}.\footnote{Note that this convention differs from the HFLAV convention by an overall minus sign in CP transformation of $D^0$, see the definitions before Eq. (131) in \cite{HeavyFlavorAveragingGroupHFLAV:2024ctg}.}

For our numerical analysis we generated, similarly to the flavor tagged sample, $N_\text{ens}^D =25$ datasets each containing $N_{\text{CP}+} = 2\times10^6$ events for the CP-even sample and the same number of $N_{\text{CP}-} = 2\times10^6$ events for the CP-odd sample. 
While these samples are substantially larger than current BESIII \cite{BESIII:2020khq} or CLEO-c \cite{CLEO:2010iul} datasets, having equal datasets in each of the three $D$ decays is a useful simplification for our exploratory study (that, however, can be easily relaxed). Like the flavor-tagged sample, each dataset was split into 90\% for training and 10\% for validation.

\paragraph{$B^\pm \to DK^\pm$ samples.}
For the likelihood fit, we generated an ensemble of $N_\text{ens}^B=15$ datasets of $B^\pm$ events following \cref{eq:hat_Gamma_minus,eq:hat_Gamma_plus} with benchmark parameters $r_B = 0.10$, $\delta_B = 130^\circ$, and $\gamma = 70^\circ$. Each dataset in the ensemble contains $N_{B^+} = N_{B^-} = 10^5$ events for each $B$ charge choice. The values of $r_B, \delta_B$ and $\gamma$ were chosen to be relatively close to the current world average~\cite{HeavyFlavorAveragingGroupHFLAV:2024ctg,HFLAV:GammaSummer2025}.
The normalization integrals, see \cref{eq:normalization} below, were evaluated with $N_{\rm MC} = 10^6$ uniformly distributed points per charge, generated independently of the fit sample.

\subsection{Unbinned likelihood for $\gamma$, $\delta_B$, and $r_B$}
\label{sec:B:likelihood}

\begin{figure}[t]
    \centering    \includegraphics[width=1.0\textwidth]{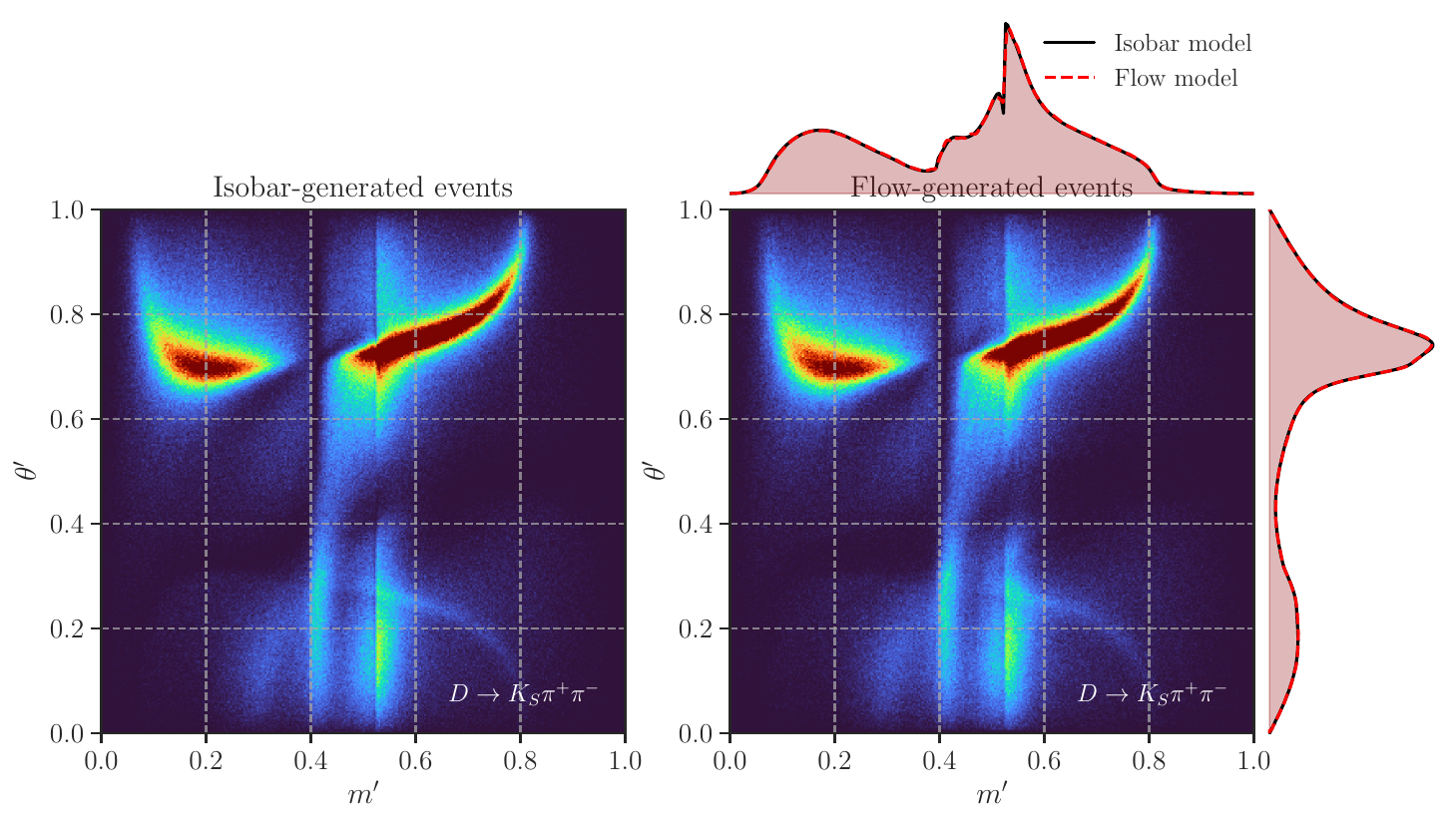}
    \caption{The $D \to K_S\pi^+\pi^-$ density in square-Dalitz coordinates for $2 \times 10^6$ events.    
    Left: isobar-generated flavor-tagged events from the first dataset in Table~\ref{tab:method_comparison} showing the resonance structure of the Dalitz plot. Right: events sampled from a normalizing flow, $K_\text{NF}(\theta',m')$, trained with the same dataset. The overlaid one-dimensional projections compare the isobar-model and flow-model marginals. 
    The $K_\text{NF}(m',\theta')$
    provides the flavor-tagged input to the unbinned $B^\pm \to DK^\pm$ likelihood.   
    }
    \label{fig:marginals}
\end{figure}

Using the continuous estimates of the $K$, $\bar K$, $\mathcal{C}$, and $\mathcal{S}$ functions, obtained in either of the two methods, we can construct the unbinned log-likelihood for the $B^\pm \to D K^\pm$ decays, that can then be used to determine $r_B, \delta_B$ and $\gamma$.
For a dataset consisting of $N_-$ $B^-$ decay events
$\{(m'_i,\theta'_i)\}_{i=1}^{N_-}$ and $N_+$ $B^+$ decay events
$\{(m'_j,\theta'_j)\}_{j=1}^{N_+}$, the negative log-likelihood is given by
\begin{equation}\label{eq:nll_full}
    -\ln \mathcal{L}(r_B, \delta_B, \gamma)
    =
    -\sum_{i=1}^{N_-} \ln f_-(m'_i, \theta'_i)
    -\sum_{j=1}^{N_+} \ln f_+(m'_j, \theta'_j)
    + N_- \ln \tilde{\mathcal{N}}_-
    + N_+ \ln \tilde{\mathcal{N}}_+\;,
\end{equation}
where $f_\pm$ are the un-normalized probability densities, cf. Eqs.~\eqref{eq:dgamma_plus_obs},
\eqref{eq:dgamma_minus_obs},
\begin{align}
    f_-(m',\theta') &= \bar{K}_\text{NF} + r_B^2 K_\text{NF} + 2 r_B \left[ \cos(\delta_B - \gamma) \, \mathcal{C}_\text{NF} - \sin(\delta_B - \gamma) \, \mathcal{S}_\text{NF} \right], \\
    f_+(m',\theta') &= K_\text{NF} + r_B^2 \bar{K}_\text{NF} + 2 r_B \left[ \cos(\delta_B + \gamma) \, \mathcal{C}_\text{NF} + \sin(\delta_B + \gamma) \, \mathcal{S}_\text{NF} \right],
\end{align}
with ${\mathcal S}_\text{NF}$ obtained from $K_\text{NF}$, ${\mathcal C}_\text{NF}$ via \cref{eq:CS_constraint},
while $\tilde{\mathcal{N}}_\pm$ are the normalization integrals 
\begin{equation}
\label{eq:N:normaliz:def}
    \tilde{\mathcal{N}}_\pm(r_B, \delta_B, \gamma) = \int_0^1 \int_0^1  f_\pm \, dm' \, d\theta'.
\end{equation}

For numerical results, the normalization factors $\tilde{\mathcal{N}}_\pm$ are evaluated via Monte Carlo integration,
\begin{equation}
\label{eq:normalization}
    \tilde{\mathcal{N}}_\pm \approx \frac{1}{N_{\rm MC}} \sum_{k=1}^{N_{\rm MC}} f_\pm(m'_k, \theta'_k),
\end{equation}
with independent samples drawn for $B^-$ and $B^+$. Since the integrand is sufficiently smooth and well-behaved over the domain, the required precision is achievable with straightforward uniform sampling, using $N_{\rm MC}=10^6$ points per $B^\pm$ charge, drawn uniformly from $[0,1]\times [0,1]$.

Note that in Eq.~\eqref{eq:nll_full} the sums are over the events in the Dalitz plot that are due to the $B^\pm\to D(\rightarrow K_S\pi^+\pi^-) K^\pm$ decays, which differ from the events in the flavor and CP-tagged $D$ decays, as would be the case in the real experiments. Since the events do not coincide,  it is thus essential to have a continuous estimate for $\tilde p_\pm$ such as via the NF based methods used in this work, to obtain the log-likelihood for $B$ decays.

\begin{figure}
    \centering
    \includegraphics[width=\linewidth]{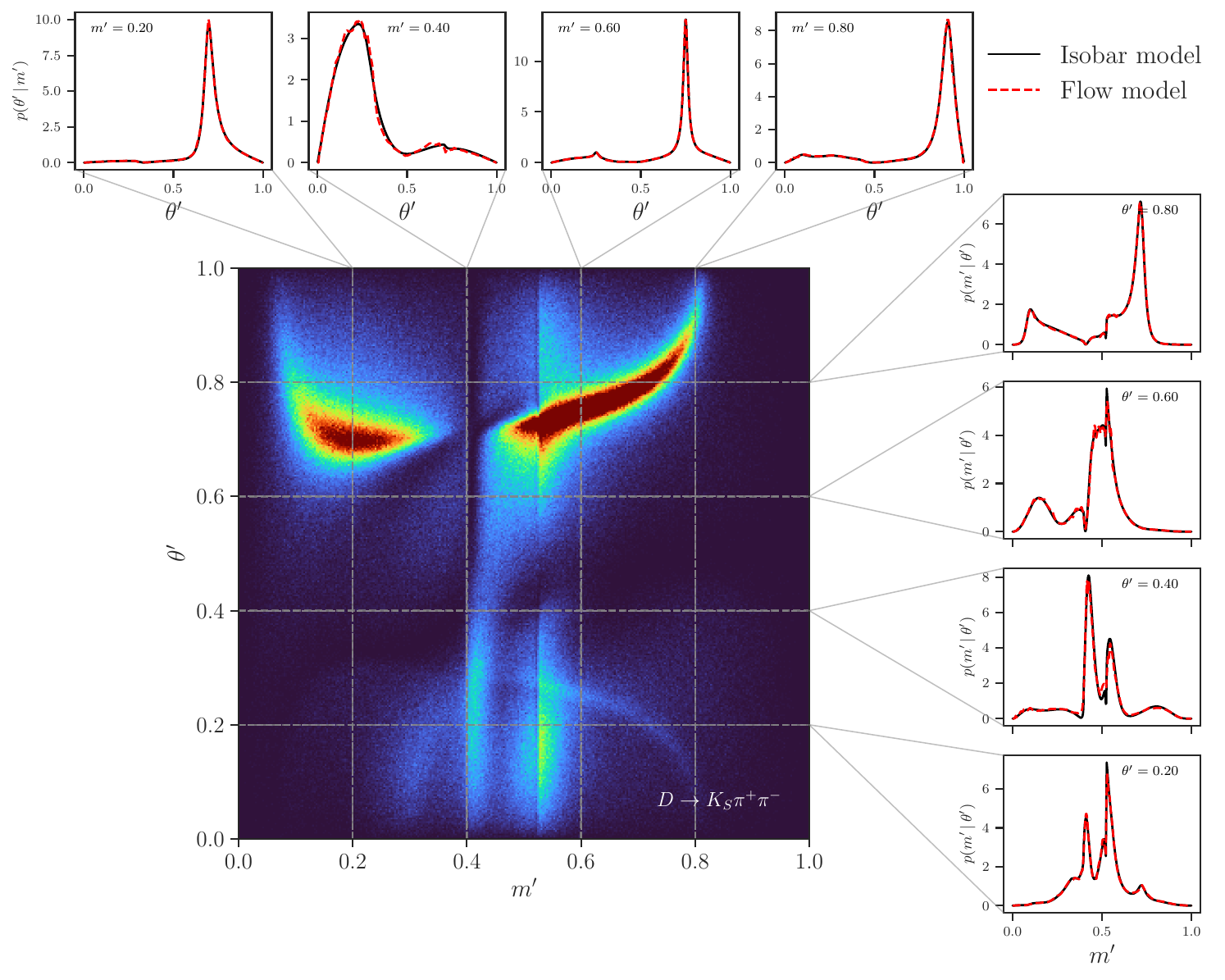}
     \caption{Conditional slices of the flavor-tagged $D\to K_S\pi^+\pi^-$ density in square Dalitz coordinates, comparing the isobar model (solid black) with the trained normalizing flow (dashed red). The central panel shows the two-dimensional flow density $K_{\mathrm{NF}}(m',\theta')$. The top row displays conditional distributions $p(\theta'\,|\,m')$ at fixed values $m'=0.20,\,0.40,\,0.60,\,0.80$, while the right column shows $p(m'\,|\,\theta')$ at fixed $\theta'=0.20,\,0.40,\,0.60,\,0.80$. The flow reproduces the narrow resonance peaks and the detailed local structure of the Dalitz density across all slices.}
    \label{fig:conditional_slices}
\end{figure}

The estimates for $(r_B, \delta_B, \gamma)$ are obtained by minimizing the negative log-likelihood in \cref{eq:nll_full}. This was performed using \textsc{Minuit} \cite{James:1975dr}, varying $r_B, \delta_B, \gamma$. In the minimization procedure we imposed the constraint that $r_B$ lies in an interval $r_B \in [0,1]$, while the two phases were allowed to vary freely,  $\delta_B, \gamma \in [0^\circ, 360^\circ)$. 
The variances of the extracted values of the parameters were obtained with the \textsc{Hesse} procedure.

\section{Numerical results}
\label{sec:numerics}

In this section, we perform a closure test: extraction of $\gamma$ from simulated data. We are specifically interested in how well the NFs reproduce the input $D\to K_S \pi^+\pi^-$ amplitude model (\cref{sec:fidelity:NF,sec:coverage}), and whether the extracted values of $r_B, \delta_B$ and $\gamma$ show any bias (\cref{sec:gamma:extract,sec:bias}).

\subsection{Fidelity of NFs}
\label{sec:fidelity:NF}

We start by first visually inspecting the fidelity of a representative NF trained on a single dataset of $N_\text{flav}=2\times 10^6$ flavor tagged $D\to K_S\pi^+\pi^-$ decays.
The comparison between the left panel (input) and the right panel (NF) in \cref{fig:marginals} shows that the learned density captures well the main resonance structures in the amplitude model. The right panel in \cref{fig:marginals} also shows that the one-dimensional marginals for the isobar model (black line) and the NF (red dashed) agree without any visible large-scale distortions.
A more detailed validation is provided in Fig.~\ref{fig:conditional_slices}, which shows conditional slices of the learned density at several fixed values of $m'$ and $\theta'$; the flow reproduces the narrow resonance peaks across the full Dalitz plot.

\begin{figure*}[t]
    \centering
    \includegraphics[width=0.95\textwidth]{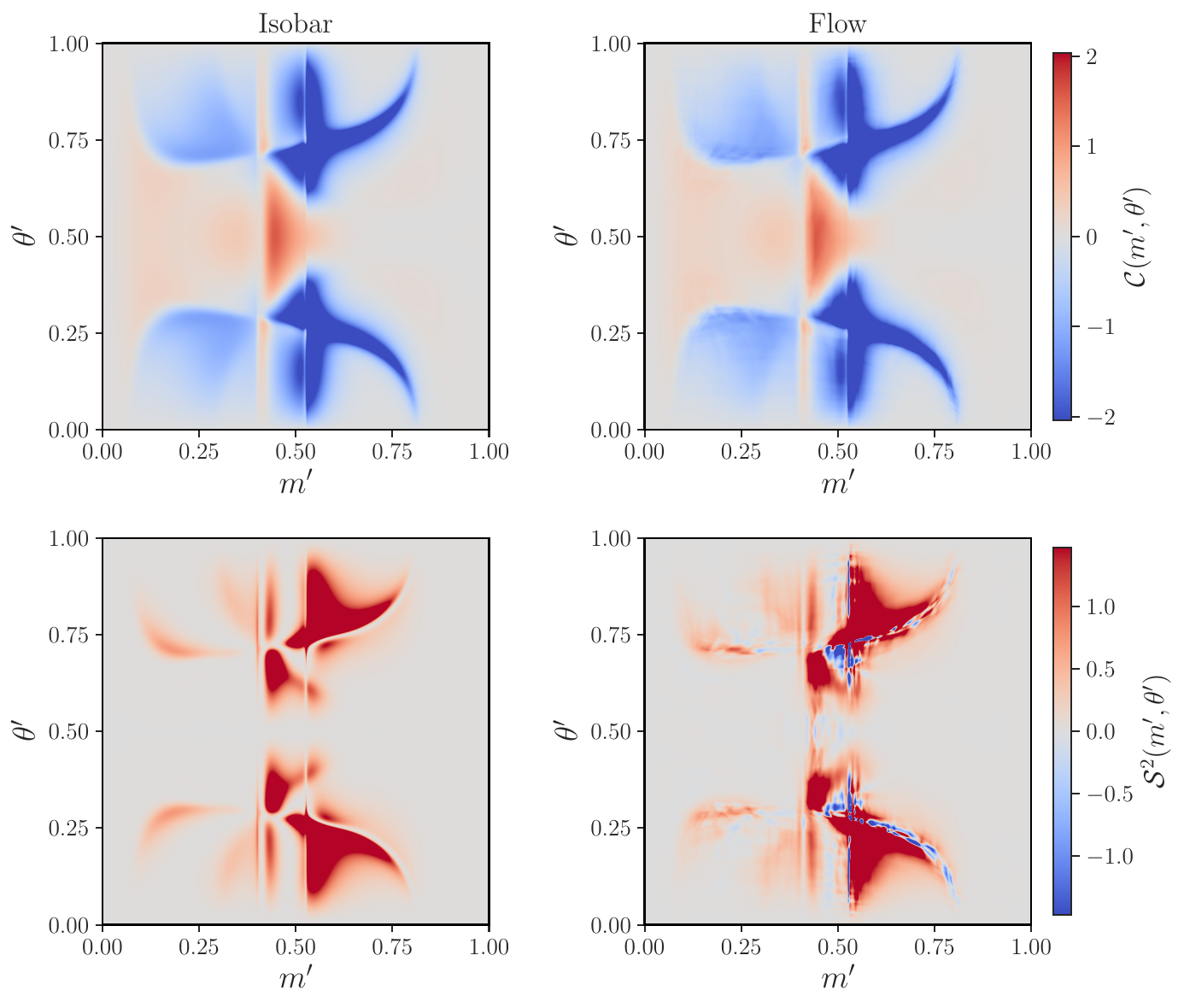}
    \caption{Comparison of the isobar-model results (the truth, 1st column) and NF-extracted results for the 3-flow approach (2nd column) for the interference observables $\mathcal{C}(m',\theta')$ (top row) and $\mathcal{S}^2(m',\theta')$ (bottom row) in square-Dalitz coordinates.
In bottom right panel the blue regions mark unphysical locations where statistical fluctuations in the independently trained NFs give $\mathcal{C}^2 > K\bar K$, so that the inferred $\mathcal{S}^2$ is negative; these points are localized near narrow resonance structures. 
}\label{fig:CS2_comparison}
\end{figure*}

\begin{figure*}[t]
    \centering
    \includegraphics[width=0.95\textwidth]{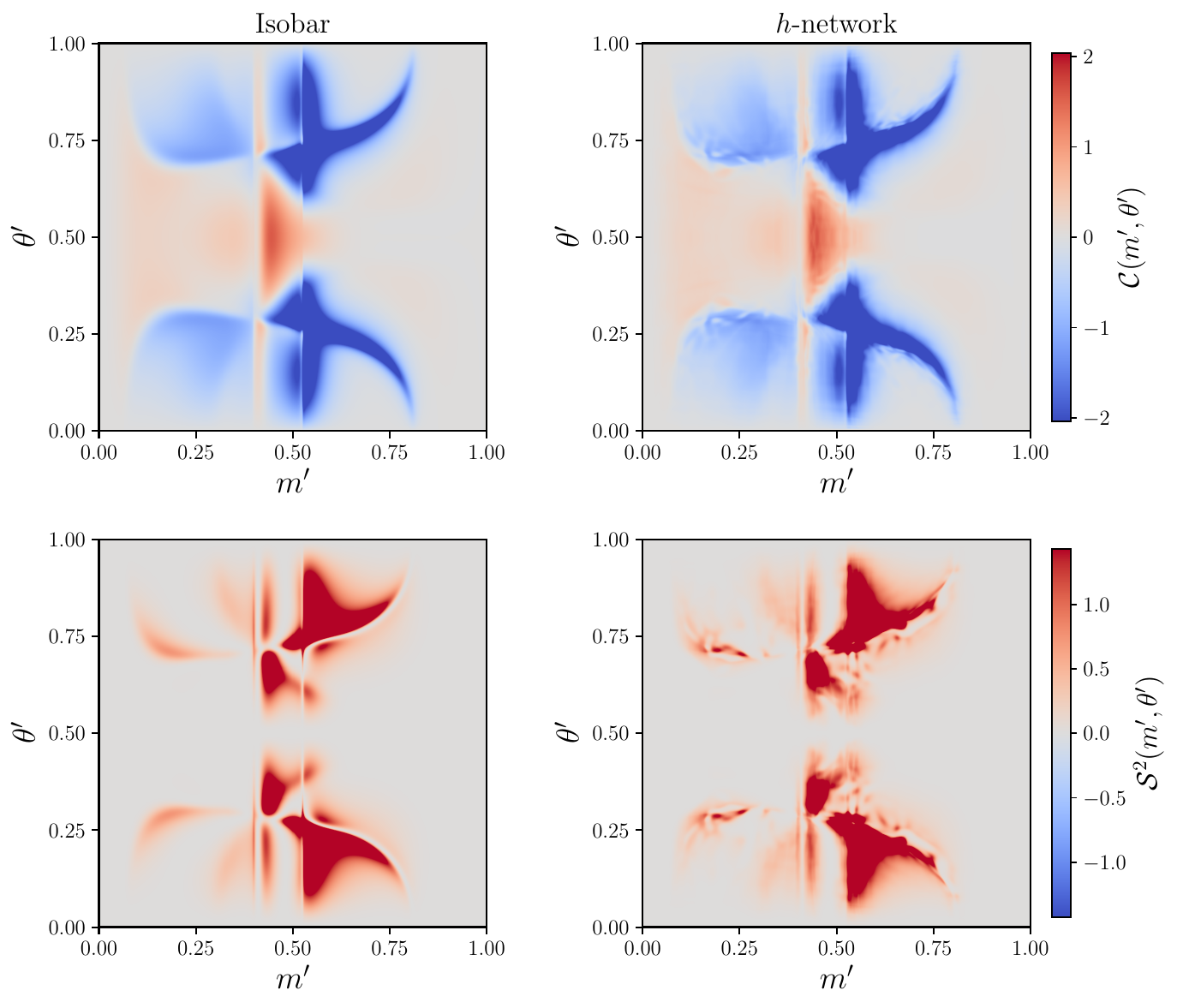}
    \caption{Same as \cref{fig:CS2_comparison} but now for the $h-$network. In this case the predicted $\mathcal{S}^2(m',\theta')$ (bottom right panel) does not reach unphysical negative values. 
    }\label{fig:CS2_comparison:h-network}
\end{figure*}

The quality of the interference information captured by the $3$-flow method NFs  $p_{\text{NF}\pm}$  trained on CP-even and CP-odd data, see \cref{eq:p+-:def}, is shown in \cref{fig:CS2_comparison}, by comparing the input and predicted values of the derived observables $\mathcal{C}(m',\theta')$ and $\mathcal{S}^2(m',\theta')$, cf.  \cref{eq:C_extraction,eq:CS_constraint}. The corresponding results for the $h-$network approach are shown in  \cref{fig:CS2_comparison:h-network}.
In both approaches  the flow-extracted $\mathcal{C}_\text{NF}$ (upper right panels in \cref{fig:CS2_comparison,fig:CS2_comparison:h-network}) follows closely the isobar-model pattern (upper left panels) across the square Dalitz plot. Despite this high-fidelity in both $K_\text{NF}$ and $\mathcal{C}_\text{NF}$, the $\mathcal{S}_\text{NF}^2$ function obtained from 
\beq
\label{eq:S2}
    \mathcal{S}_\text{NF}^2(m',\theta') = K_\text{NF}(m',\theta') \bar{K}_\text{NF}(m',\theta')- \mathcal{C}_\text{NF}^2(m',\theta'),
\end{equation}
takes unphysical values in the $3-$flow case in a small region close to the peaks  of the resonances, where $\mathcal{C}^2 > K\bar K$. These are visible as blue regions in the bottom right panel in \cref{fig:CS2_comparison}. The negative $\mathcal{S}_\text{NF}^2$ regions are concentrated near the narrow resonance structures, where the amplitudes vary most rapidly and the independent density estimates are most sensitive to statistical fluctuations. In comparison, the isobar model (bottom left panel) shows no negative values for $\mathcal{S}^2$, as expected for a truth level model. The $h-$network predictions for $\mathcal{S}_\text{NF}^2$ (bottom right panel in \cref{fig:CS2_comparison:h-network}), similarly does not predict unphysical negative $\mathcal{S}_\text{NF}^2$, since the constraint \cref{eq:constraint:C} is now built-in into the ML architecture via $|h_\text{MLP}|\leq 1$, cf. \cref{eq:h:definition}. In contrast to small $\mathcal{S}_\text{NF}^2$ value regions, the extracted $\mathcal{S}_\text{NF}^2$ can never be ``too big'', i.e.,  one always has $\mathcal{S}^2 \leq K\bar K$, which follows immediately from the definition in  \cref{eq:S2}.

As we will show below, the appearance of unphysical values for ${\mathcal S}_\text{NF}^2$ in the $3$-flow method is not merely an annoyance, but rather has practical consequences. Since in this case one needs to correct a posteriori the predictions of NFs by an averaging procedure to obtain physical results, this necessarily leads to some statistical loss in the sensitivity to $\gamma$ \cite{Bondar:2008hh,Bondar:2005ki}. In the numerics we use the local neighbor-averaging procedure described in \cref{sec:unbinned_fit}., but one could also use alternative approaches, such, as for instance, simply disregard $B$ decay events with negative ${\mathcal S}_\text{NF}^2$ values (which clearly would lead to statistical loss in sensitivity to $\gamma$). 

\subsection{Coverage studies}
\label{sec:coverage}

\begin{figure*}[t]
    \centering
    \includegraphics[width=0.95\textwidth]{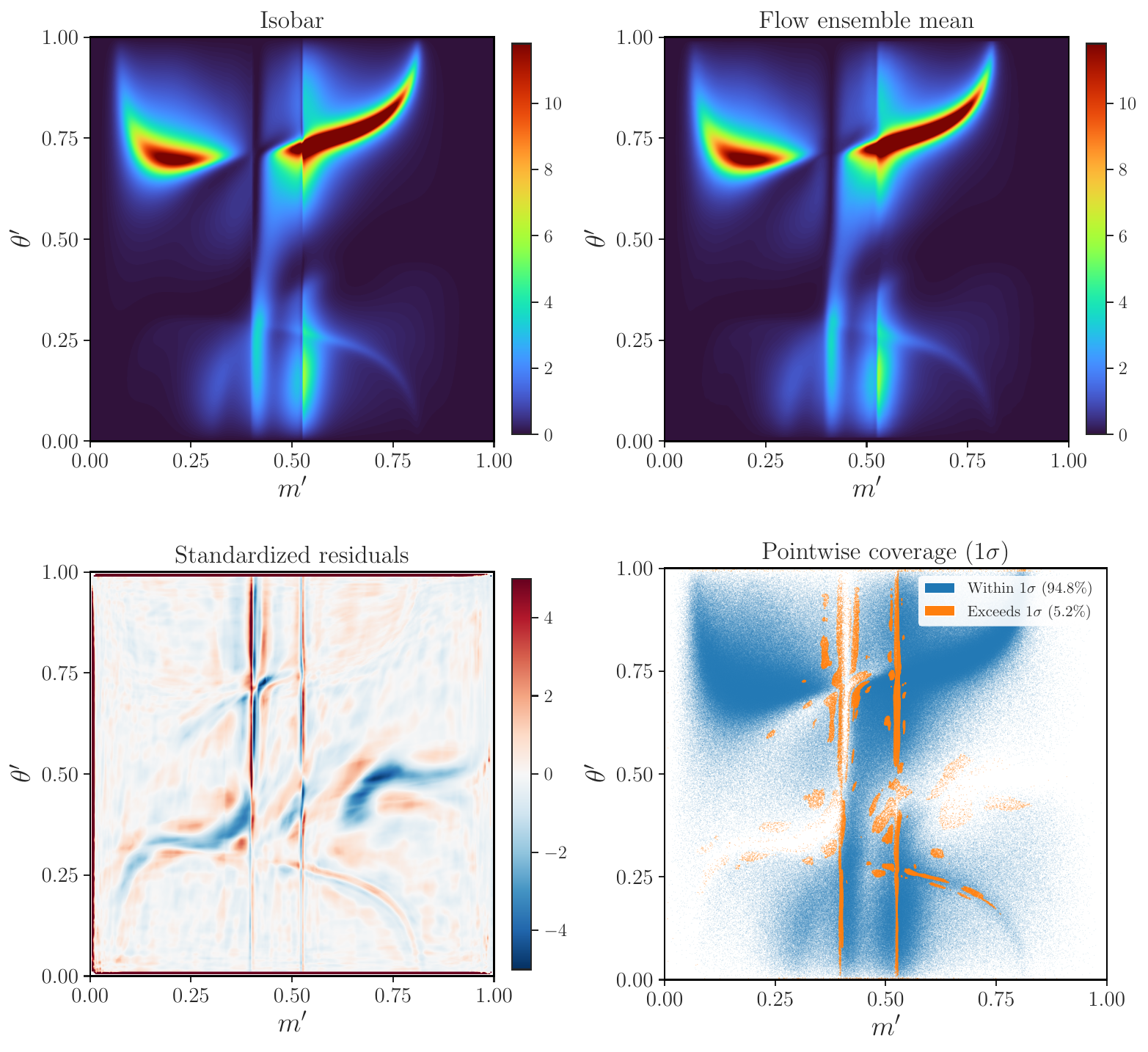}
    \caption{Coverage diagnostics for the flavor tagged NF ensemble of $N_\text{ens}^D=25$ independently trained models. {\em Upper left:}~Isobar-model density $\rho_{\rm iso}(m',\theta')$. {\em Upper right:}~Flow ensemble mean density $\langle\rho(m',\theta')\rangle$. {\em Bottom left:}~Standardized residuals $(\langle\rho\rangle - \rho_{\rm iso})/\sigma_{\rm ens}$; deviations beyond $\pm 2$ are concentrated along narrow resonance bands. {\em Bottom right:}~Pointwise $1\sigma$ coverage map: within $1\sigma$ (blue, $94.8\%$), and exceeding $1\sigma$ (orange, $5.2\%$).}
    \label{fig:coverage_test}
\end{figure*}

To further quantify the confidence in the learned NF densities for the $3-$flow method, we examine the point-wise coverage across the square $D\to K_S \pi^+\pi^-$ Dalitz plot using $N_\text{ens}^D=25$ independent datasets (further studies of the learned neural-network fidelity for the $h$-network approach can be found in Appendix~\ref{app:constraint_aware}). At each square-Dalitz point $z_k = (m'_k, \theta'_k)$, we compute the ensemble mean and standard deviation, 
\begin{equation}
  \langle \rho(z_k) \rangle = \frac{1}{N_{\rm ens}^D}\sum_{n=1}^{N_{\rm ens}^D}\rho_n(z_k),
  \qquad
  \sigma_{\rm ens}(z_k) = \sqrt{\frac{1}{N_{\rm ens}^D-1}\sum_{n=1}^{N_{\rm ens}^D}\bigl(\rho_n(z_k)-\langle \rho(z_k) \rangle\bigr)^2},
\end{equation}
where $\rho$ is any of the three NFs, $\rho=\{K_\text{NF}, p_{\text{NF}\pm}\}$.
A comparison of $\langle \rho \rangle$ for $K_\text{NF}$ with the isobar-model density $\rho_{\rm iso}$ is shown in the top two panels in \cref{fig:coverage_test}. The ensemble mean $\langle K_\text{NF} \rangle$  reproduces the resonance structure well across the full phase space. Similar comparisons of ensemble means with isobar model densities for $p_{\text{NF}\pm}$ are shown in Appendix \ref{app:cp_coverage}, showing similar high fidelity.

We also define a point-wise $1\sigma$ coverage fraction $f_\text{cov}$, i.e., a fraction of the $D$ decays at which the disagreement with the isobar model is more than one sigma deviation over the ensemble.  Practically, this is calculated using $N_\text{grid}$ square Dalitz points that are distributed according to the isobar model density, with $f_\text{cov}$ given by 
\begin{equation}
  f_{\rm cov} = \frac{1}{N_{\rm grid}}\sum_{k=1}^{N_{\rm grid}}
  \Theta\left[|\rho_{\rm iso}(z_k) - \langle \rho(z_k)\rangle| - \sigma_{\rm ens}(z_k)\right],
\end{equation}where $\Theta$ is the Heaviside 
function. 
We take $N_\text{grid}=N_{\rm flav} = 2 \times 10^6$.

 Over the ensemble of $N_\text{ens}^D=25$ $D$ decay datasets we find $f_{\rm cov} = 94.8\%$, with the
corresponding points denoted as blue regions in the bottom-right panel in \cref{fig:coverage_test}. The remaining $5.2\%$ of points, shown in orange, which lie outside the expected $1\sigma$ error band around the isobar input, are concentrated along narrow resonance bands where the amplitude varies most rapidly. As we will see in the next two subsections, these regions are small enough that they carry only relatively small statistical weight in the $B^\pm \to DK^\pm$ likelihood. The standardized residual ($(\langle \rho(m',\theta')\rangle - \rho_{\rm iso}(m',\theta'))/\sigma_{\rm ens}(m',\theta')$) map confirms that deviations beyond $\pm 2\sigma_\text{ens}$ are likewise localized to these narrow structures, as seen in the bottom left panel in  \cref{fig:coverage_test}. Analogous diagnostics for the CP-even and CP-odd flows yield qualitatively similar results, with $1\sigma$ coverage fractions of $97.5\%$ and $96.0\%$, respectively (see App.~\ref{app:cp_coverage} for details).

\subsection{Extraction of \texorpdfstring{$\gamma$}{gamma}}
\label{sec:gamma:extract}

\begin{table}[t]
    \centering
    \begin{tabular}{clccc}
        \hline\hline
        Dataset & Method & $r_B$ & $\delta_B\,[^\circ]$ & $\gamma\,[^\circ]$ \\
        \hline
        \multirow{2}{*}{1}
        & 3-flow     & $0.0986 \pm 0.0017$ & $130.63 \pm 0.87$ & $69.92 \pm 0.89$ \\
        & $h$-network & $0.1004 \pm 0.0017$ & $128.68 \pm 1.10$ & $67.65 \pm 1.10$ \\[4pt]
        \vdots &  \vdots & \vdots & \vdots & \vdots  \\[4pt]
        \multirow{2}{*}{3}
        & 3-flow     & $0.1003 \pm 0.0017$ & $130.30 \pm 1.08$ & $68.34 \pm 1.09$ \\
        & $h$-network & $0.1011 \pm 0.0017$ & $129.44 \pm 1.09$ & $68.73 \pm 1.09$ \\[4pt]
        \vdots &  \vdots & \vdots & \vdots & \vdots  \\[4pt]
        \multirow{2}{*}{5}
        & 3-flow     & $0.0990 \pm 0.0017$ & $130.24 \pm 1.11$ & $67.04 \pm 1.12$ \\
        & $h$-network & $0.0986 \pm 0.0017$ & $129.24 \pm 1.10$ & $70.04 \pm 1.10$ \\[4pt]
        \vdots &  \vdots & \vdots & \vdots & \vdots  \\
        \multirow{2}{*}{Avg.}
        & 3-flow     & $0.0989 \pm 0.0029$ & $129.91 \pm 1.56$ & $68.91 \pm 1.82$ \\
        & $h$-network & $0.1001 \pm 0.0018$ & $128.86 \pm 1.20$ & $68.43 \pm 1.43$ \\
        \hline
        --- & Amp.\ model & $0.1010 \pm 0.0017$ & $129.40 \pm 1.10$ & $69.34 \pm 1.10$ \\
        \hline
        --- & Benchmark   & $0.1000$ & $130.00$ & $70.00$ \\
        \hline\hline
    \end{tabular}
    \caption{Fitted values of $r_B$, $\delta_B$, and $\gamma$ for three representative datasets, comparing the constraint deferred (three-flow) and constraint-aware ($h$-network) methods. The amplitude-model fit uses interference observables from the isobar model on the same $B^\pm$ dataset, isolating the irreducible statistical uncertainty. For individual datasets, the uncertainties quoted are the $1\sigma$ Hesse errors from the fit, $\sigma_\text{i}^\text{Hesse}$, while for averages $\sigma_\text{tot}$ is quoted, cf. \cref{eq:bar:x,eq:sigma:tot,eq:sigma:flow,eq:sigma:fit} in the main text.
    }
    \label{tab:method_comparison}
\end{table}

\begin{figure}[t]
    \centering
    \includegraphics[width=0.95\linewidth]{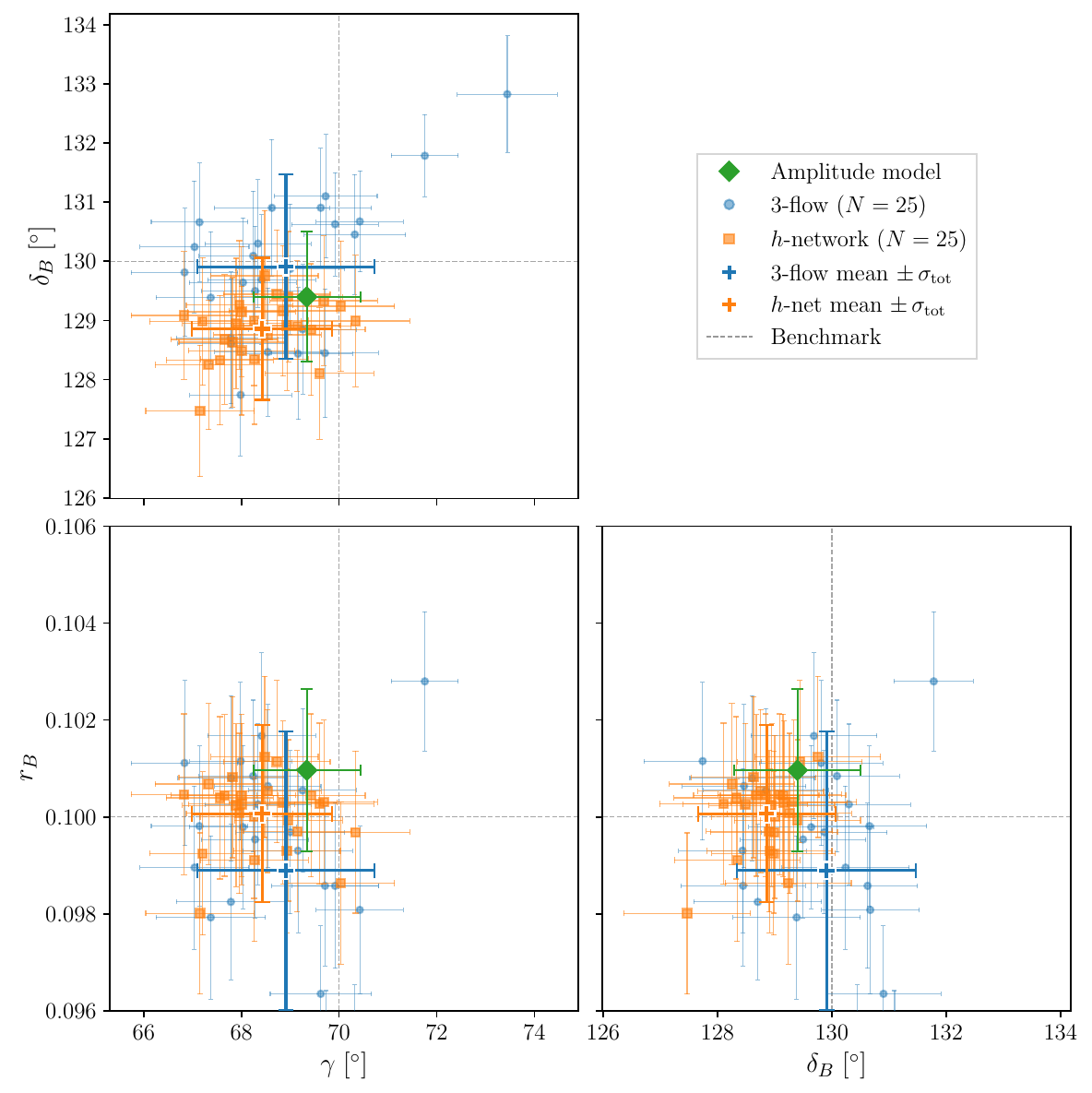}
    \caption{Corner plot comparing the three-flow (blue circles) and constraint-aware $h$-network (orange squares) methods across 25 shared datasets. Each point shows one fit with its $1\sigma$ uncertainty obtained from Minuit.
    The larger cross-hair markers indicate the method-specific means with total uncertainty $\sigma_{\rm tot} = \sqrt{\sigma_{\rm flow}^2 + \sigma_{\rm fit}^2}$. The green diamond is the amplitude-model fit and the dashed lines mark the benchmark values.
    }
    \label{fig:corner_comparison}
\end{figure}

We compare the two extraction methods on a common footing: the constraint-aware method that uses $K_\text{NF}$ and $h_\text{MLP}$ neural networks, and the deferred  constraint enforcement method that uses three NFs, $K_\text{NF}$, $p_{\text{NF}\pm}$ together with the averaging procedure in the regions with unphysical estimates for the derived quantity ${\mathcal  S}_\text{NF}^2$.
 All $N_\text{ens}^D=25$  fits use the same fixed $B^\pm \to DK^\pm$ pseudo-data ($N_{B^+} = N_{B^-} = 10^5$),  with the negative log-likelihood given in \cref{eq:nll_full}, but different realizations of $D$ decay data (the effect of varying $B$ decay data will be explored in the next subsection as well).
 The differences in the extracted values of $r_B$, $\delta_B$, and $\gamma$ therefore arise solely from the statistical fluctuations in $D\to K_S\pi^+\pi^-$ data, which then translate to differences in NF density estimations of  ${\mathcal C}_\text{NF}$ and ${\mathcal S}_\text{NF}$ for all points, and from the treatment of ${K}_\text{NF}$ at points that would otherwise yield unphysical values of $\mathcal{S}_\text{NF}^2$.
As a reference, we also perform an ``amplitude-model fit'' in which $K$, $\mathcal{C}$, and $\mathcal{S}$ are taken directly from the isobar model rather than from any learned density. 

\Cref{tab:method_comparison} shows the fitted parameters for three representative datasets alongside the fits obtained using the truth level amplitude model for reference. We see that the extracted values of $r_B, \delta_B$ and $\gamma$ mostly agree with the input values within estimated statistical uncertainties.
\Cref{fig:corner_comparison} presents the full  $N_\text{ens}^D$ dataset comparison as a corner plot, displaying the two-dimensional correlations between all parameter pairs. Each point carries its per-fit Hesse uncertainty, while the larger cross-hair markers 
indicate the unweighted ensemble mean of the fitted central values,
\beq
\label{eq:bar:x}
  \bar{x}
  =
  \frac{1}{N_\text{ens}^D}
  \sum_{i=1}^{N_\text{ens}^D} x_i ,
\eeq
together with the total uncertainty
\beq
\label{eq:sigma:tot}
  \sigma_{\rm tot}
  =
  \left(\sigma_{\rm fit}^2+\sigma_{\rm flow}^2\right)^{1/2}.
\eeq
Here,
\beq
\label{eq:sigma:flow}
  \sigma_{\rm flow}^2
  =
  \frac{1}{N_\text{ens}^D-1}
  \sum_{i=1}^{N_\text{ens}^D}
  \left(x_i-\bar{x}\right)^2,
\eeq
is the sample variance of the fitted central values across the ensemble, while
\beq
\label{eq:sigma:fit}
  \sigma_{\rm fit}
  =
  \frac{1}{N_\text{ens}^D}
  \sum_{i=1}^{N_\text{ens}^D}
  \sigma_i^{\rm Hesse},
\eeq
is the arithmetic mean of the per-fit Hesse uncertainties. That is, $\sigma_\text{flow}$ 
quantifies the spread induced
by the normalizing-flow across different $D$ decay datasets,  while the $B$ decay dataset, which is the source of per-fit Hesse uncertainty $\sigma_\text{fit}$, was held fixed. An improved estimate of $\sigma_\text{fit}$ is thus obtained via an unweighted ensemble average.

The results for the mean and $\sigma_\text{tot}$ for each of the extracted parameters, $r_B, \delta_B$ and $\gamma$ are also listed in \cref{tab:method_comparison}, in the ``Avg.'' row.
Note that the first dataset in \cref{tab:method_comparison} is also the one that was used to illustrate the fidelity of NFs in \cref{sec:fidelity:NF}.
Two features stand out from the comparisons in \cref{tab:method_comparison} and \cref{fig:corner_comparison}. First, the $h$-network extracts $r_B$ with smaller variance across datasets than the three-flow method.  This is a direct consequence of the constraint $|h|\leq 1$, which  by construction guarantees $\mathcal{C}^2 \leq K\bar{K}$ in the $h$-network approach. In the three-flow approach, constraint violations require numerical repair via nearest neighbor averaging, which can distort the learned interference amplitude.

\subsection{Study of potential biases}
\label{sec:bias}

\begin{figure}[t]
    \centering
    \includegraphics[width=\textwidth]{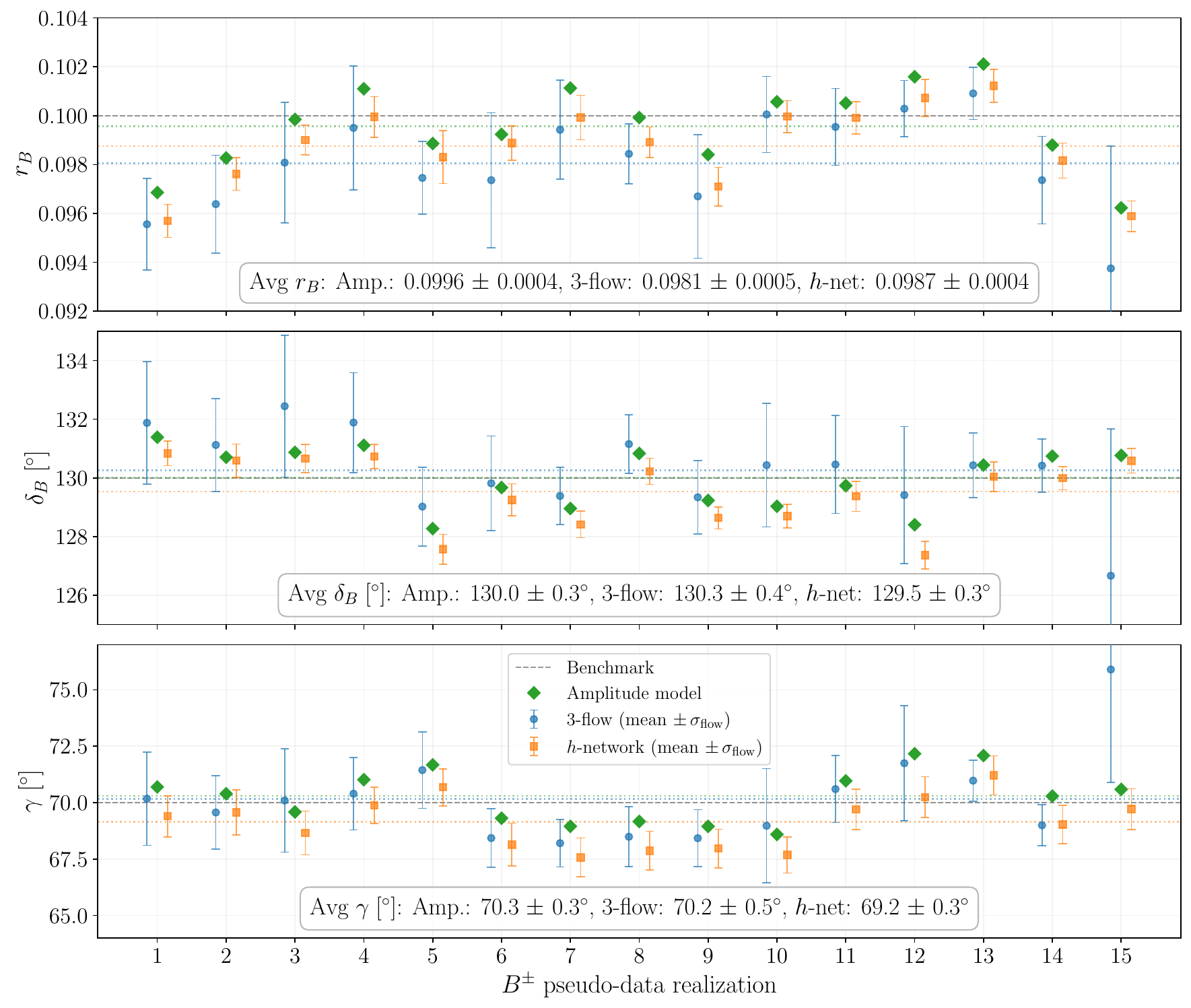}
    \caption{Extracted values of $r_B$ (top), $\delta_B$ (middle), and $\gamma$ (bottom) across 15 independent $B^\pm$ pseudo-data realizations. Green diamonds show the amplitude-model fits, blue circles the three-flow method, and orange squares the $h$-network. Error bars indicate $\sigma_{\rm flow}$, the spread over the 25 $D$ decay ensembles. Dashed lines mark the averages, with each color corresponding to the methods. The flow-based results track the amplitude-model scatter closely, with hints of small residual biases discussed in the text.
    }
    \label{fig:bias_study}
\end{figure}
 
The results in the previous subsection were obtained using a single $B^\pm$ pseudo-data sample. While the spread across the 25 $D$ decay ensembles provides a measure of the density-estimation uncertainty, it does not fully capture potential biases in the extracted values of $\gamma$, $\delta_B$, or $r_B$, i.e., a potential systematic shift common to all ensemble members. To properly assess whether the flow-based extraction introduces a bias on $\gamma$, $\delta_B$, or $r_B$, we must vary the $B$ decay data as well.
 
To this end, we repeat the full extraction pipeline across $N_\text{ens}^B=15$ independent $B^\pm$ pseudo-data realizations, each generated with the same benchmark parameters ($r_B = 0.10$, $\delta_B = 130^\circ$, $\gamma = 70^\circ$) but with independent statistical fluctuations. For each $B^\pm$ decay realization we perform $r_B, \delta_B, \gamma$ extraction for $N_\text{ens}^D=25$ datasets of $D$ decays. From this we obtain the mean value of extracted parameters, as well as $\sigma_\text{tot}$, in the same way as discussed in detail for a single $B^\pm$ realization in previous subsection. 
The amplitude-model fit is performed in parallel on each $B$ dataset as a bias-free reference.
 
The results are shown in \cref{fig:bias_study}. For each of the three parameters, the amplitude-model fits (green diamonds) scatter around the injected values as expected from uncorrelated $B$ sample fluctuations. The mean across $N_\text{ens}^B$ $B$ decay realizations (green dotted line) is very close to the injected value of the parameter (grey dashed line). 
The flow-based methods track the scatter in the amplitude fit extractions closely, indicating that the density-estimation step does not introduce large distortions, which would vary from one $B$ dataset to another. Note that the uncertainties on extracted values of parameters using the $h-$network approach are stable across different $B$ datasets. In contrast, the nearest neighbor averaging required in the  3 flow method introduces a larger variability of uncertainties on $r_B, \delta_B$ and $\gamma$ across the $B$ decay datasets. For $B$ decay dataset No. 15, in particular, the fit for one out of 25 $D$ decay  datasets converges to $r_B$, $\delta_B$, $\gamma$ values that are far from the benchmark input values, resulting in overall large error bars.
 
The spread between the values of parameters extracted using the truth-level isobar model and the NF based extractions is below $1\sigma$ statistical error for virtually all examples in \cref{fig:bias_study}, indicating that there is no statistical evidence for a systematic bias. 
 In order to obtain an upper bound on the possible size of the bias we proceed by averaging the results over the $N_\text{ens}^B=15$ realizations, treating the $D$ decay datasets as independent (even though they are fully correlated across $B$ decay realizations). The amplitude-model fits yield $r_B = 0.0996(4)$, $\delta_B = 130.0(3)^\circ$, and $\gamma = 70.3(3)^\circ$, all consistent with the benchmark input values within $1\sigma$ as expected, since the errors in this case have the correct statistical interpretation. The three-flow method gives $r_B = 0.0981(5)$, $\delta_B = 130.3(4)^\circ$, and $\gamma = 70.2(5)^\circ$, while the $h$-network yields $r_B = 0.0987(4)$, $\delta_B = 129.5(3)^\circ$, and $\gamma = 69.2(3)^\circ$. Note that in this case the quoted errors do not have a clear statistical interpretation due to use of fully correlated $D$ decay samples (they are quoted merely as a consistency check with the isobar model results). 
 We view the differences of the averages to the inputs as the upper bounds on the possible biases on the parameter extractions using NFs. The largest one are, for the $h$-network, 
a $\sim 1^\circ$ downward shift in $\gamma$, and  $\sim -0.5^\circ$ shift in  $\delta_B$, and for 
three-flow method  a  $\sim 0.002$ 
downward shift in $r_B$.
Each of these shifts is smaller or at the level of per-fit statistical uncertainties, which are $\sim 0.002-0.003$ for $r_B$ and $\sim 1^\circ-2^\circ$ for $\delta_B$ and $\gamma$.

These results are encouraging: the upper bounds on biases are small relative to current experimental uncertainties on $\gamma$, and both methods successfully recover the injected parameters within their combined uncertainties. Nevertheless, a more detailed investigation is warranted, including studies with varied $D$ sample sizes and flow architectures, to fully characterize and ultimately reduce any potential biases. We discuss this further in \cref{sec:Implementations}.

\section{Toward implementation with real data}
\label{sec:Implementations}
While our study shows that the NF methods can enable a clean extraction of the parameters, implementing them in a real experimental analysis requires further work.

One important aspect is error propagation. 
Our MC study did not include experimental uncertainties, such as those on final-state particle energies or particle identification. 
These effects should be incorporated in any realistic analysis, and can in principle be embedded directly within the NF framework, providing a unified treatment. Because NFs provide a local density estimate across the Dalitz plot, systematic uncertainties can be applied point-wise rather than averaged over bins; the estimates from current binned analyses can be used already as the starting point.

Another issue are potential biases. The multi-$B$-dataset study in Sec.~\ref{sec:bias} did not reveal any statistically significant biases in the extracted values of the parameters, with estimated upper bounds on possible residual biases at the level of one degree in $\gamma$ and $\delta_B$ and at per-mille level on $r_B$.
While such potential shifts are small compared to the current experimental uncertainties, they could become relevant as $B$ decay samples grow at LHCb and Belle~II. Developing robust methods to quantify and correct for such biases, such as experimenting with more expressive flow architectures, will be an important step toward a real-data implementation.

It is also crucial to understand the overall statistical uncertainties and how they propagate through the analysis chain. The unbinned likelihood for $\gamma$, $\delta_B$, and $r_B$ depends on the learned functions $K$, $\mathcal{C}$, and $\mathcal{S}$ at every point in the Dalitz plot, so uncertainties in these density estimates propagate directly into the fitted parameters. In scenarios where the $B$ data is the limiting factor, the uncertainty associated with the NF description of the $D$ data should be negligible, since the per-point density errors are suppressed by the large $D$ sample size. While there is significantly more $D^*$-tagged data than $B$ data, this is not necessarily the case for the quantum-correlated $D$ samples from $\psi(3770) \to D\bar{D}$ that determine $\mathcal{C}$ and $\mathcal{S}$, which are expected to remain limited for the foreseeable future. In such cases, a careful study of how the statistical uncertainties in $\mathcal{C}$ and $\mathcal{S}$ propagate into the errors on $\gamma$ is required. One approach is to set aside part of the data for training and use the remainder for the analysis, at the cost of a small loss in statistical power, as was done in this paper. However, a more principled treatment of this uncertainty could be obtained by replacing the deterministic flows with Bayesian normalizing flows~\cite{mackay1995probable,neal1995bayesian,gal2016uncertainty,Louizos2017mnf,Bollweg:2019skg,Ernst:2023qvn,Bierlich:2023zzd,Butter:2025wxn}, in which a posterior distribution over the network weights is maintained throughout training. This would allow the uncertainties in $K$, $\mathcal{C}$, and $\mathcal{S}$ to be automatically propagated into the $\gamma$ extraction via marginalization over the weight posterior, without the need for an external ensemble or data-splitting procedure.

It is also important to compare the NF method with the by-now standard binned approach \cite{Giri:2003ty,Bondar:2008hh}. In the latter case, statistical uncertainties are well understood, and the main loss in sensitivity arises from binning. With the current 8-bin scheme, this dilution is at the level of about $10\%$. With more correlated $D$ data, one could consider finer binning, potentially reducing this loss, although a dedicated study of the achievable improvement is still lacking.

Comparing the two methods, we find that they are subject to different types of uncertainties. We therefore advocate using them side by side, both as a cross-check and as a way to identify potential biases and unaccounted-for systematic effects.

\section{Conclusions}
\label{sec:conclusions}

We introduced an unbinned method for extracting the CKM angle $\gamma$ from $B^\pm \to (D \to K_S\pi^+\pi^-)K^\pm$ decays using normalizing flows as continuous density estimators of the $D$ decay amplitude over the square Dalitz plot. The key advantage over the standard model-independent binned approach is that the NF representation improves systematically with larger $D$ decay samples, in direct analogy with how the $c_i$ and $s_i$ parameters in the BPGGSZ method benefit from larger charm-threshold datasets.

A central technical challenge is that the three functions entering the $B^\pm$ likelihood, $K$, $\mathcal{C}$, and $\mathcal{S}$, are not independent but rather satisfy the trigonometric constraint $\mathcal{C}^2 + \mathcal{S}^2 = K\bar{K}$. We explored two strategies for handling this. In the three-flow (deferred constraint) approach, the flavor-tagged and CP-tagged densities are learned independently, and violations of the constraint are repaired by local neighbor averaging. In the constraint-aware ($h$-network) approach, a SIREN network $h_\text{MLP}$ with $|h|\leq 1$ is used to parameterize $\cos\Delta\delta$ directly, so the constraint is satisfied exactly at every Dalitz point by construction.

In a closure test based on $N_\text{ens}^D = 25$ independently generated flavor-tagged and CP-tagged $D$ decay datasets of $2\times 10^6$ events each, and $N_\text{ens}^B$=15 independently generated $B^\pm$ decay datasets with $N_{B^\pm} = 10^5$ signal events per charge, both methods recover the injected parameters $(r_B, \delta_B, \gamma) = (0.10, 130^\circ, 70^\circ)$ within uncertainties. The $h$-network yields smaller variance in $r_B, \delta_B$ and $\gamma$ across datasets, a direct benefit of the built-in constraint.
We find that the proposed architectures do not lead to any statistically significant biases in the extracted values of parameters, and estimate that they are smaller or at worst comparable to the statistical uncertainties in the simulated datasets. In particular, they are much smaller than the current experimental uncertainties, but may need to be controlled as LHCb and Belle~II accumulate larger $B$ decay samples, for instance, by considering deeper/wider networks, or by exploring other architectures equally well-suited to modeling the multi-modal target functions encountered in  the $\gamma$ fit.

The present study is a proof of principle. Several ingredients needed for a full experimental implementation remain to be addressed: detector acceptances, backgrounds, and experimental resolution.
On the ML architecture side, replacing the deterministic flows with Bayesian normalizing flows would provide per-point uncertainty estimates on $K$, $\mathcal{C}$, and $\mathcal{S}$, allowing the modeling uncertainty to be propagated into $\gamma$ via marginalization rather than through an external ensemble. 

It is useful to compare the proposed NF based $\gamma$ extraction with the traditional model independent binned method. In the binned method, there are essentially no theoretical errors, however, binning does lead to a reduction in statistical power. Unbinned methods, in contrast, aim to minimize the statistical uncertainty, but usually at the cost of introducing some theoretical errors. In the approach introduced in this paper, the theory uncertainties are related to the precision with which the NF models the Dalitz plot. The key novelty of using NFs is their ability to model complex distributions with many parameters. An extension to the Bayesian NFs would then also capture the statistical and modeling uncertainties {\em automatically}. 
Given this complementarity between the traditional binned $\gamma$ extractions and the unbinned extractions that would utilize advances in machine learning for density estimations, we argue that both options should be pursued by the experiments. We regard the results presented here as a concrete starting point for such a program.

\section*{Acknowledgments}
We thank Marat Freytsis for discussions at the initial stages of this project, and for help with the amplitude code. We thank Phil Ilten for discussions. YG is supported by the NSF grant PHY-2309456. TM is supported in part by the Shelby Endowment for Distinguished Faculty at the University of Alabama and by Fermilab via Subcontract 725339. JZ acknowledges support in part by DOE
grants DE-SC0011784 and DE-SC0026301, and by NSF grants OAC-2103889, OAC-2411215, and OAC-2417682. S.S.~is supported by the STFC through an Ernest Rutherford Fellowship under reference ST/Z510233/1 and the grant ST/X003167/1. Fermilab is managed by Fermi-Forward Discovery Group, LLC, acting under Contract No.~89243024CSC000002 with the U.S. Department of Energy, Office of Science, Office of High Energy Physics.

\begin{appendix}

\section{Normalizing flows: architecture and training}
\label{app:nfs}
Normalizing flows (NFs)~\cite{dinh2015, Kobyzev_2021, Rezende:2015a} are generative models that represent a complicated target density as the image of a simple latent density under an invertible map. In our application they provide flexible two-dimensional density estimators with exact likelihood evaluation.

Given a random variable $\bz \in \mathbb{R}^d$ and an invertible map $f : \mathbb{R}^d \to \mathbb{R}^d$, the probability density of $\bx = f(\bz)$ is
\begin{equation}
  \lhp[X,f](\bx) = \lhp[Z](\bz)|\det J_{f} (\bz)|^{-1} \mmp{,}
  \label{eq:prob_transform}
\end{equation} 
where $J_{f} = \partial f / \partial \bz$ is the Jacobian of the transformation. The full NF map is a composition of $n$ such transformations, $\bz \equiv \bz_0 \to \bz_1 \equiv f_1(\bz_0) \to \cdots \to \bx \equiv \bz_n$, so that
\begin{equation}
  \lhp[X](\bx) = \lhp[Z](\bz) \prod_{i = 1}^n
  |\det J_{f_i} (\bz_{i - 1})|^{-1} \mmp{.}
\end{equation}

The model parameters are determined by minimizing the negative log-likelihood on the training sample. Given events $\bx_a = \{\bx_1, \bx_2, \ldots, \bx_N\}$ with optional conditional labels $\bc_a=\{\bc_1,\bc_2,\dots,\bc_N\}$, the loss is
\begin{equation}
  \begin{aligned}
    \llh &= \mathbb{E}_{\lhp[X](\bx,\bc)}
    \left[ - \log \lhp[X](\bx;\btheta, \bc)\right] = -\frac{1}{N}\sum_{a= 1}^N
    \log \lhp[X](\bx_a;\btheta, \bc_a) \\
    &=  \frac{1}{N}\sum_{a = 1}^N \left\{ - \log \lhp[Z]
    \left(F^{-1}(\bx_a; \btheta, \bc_a)\right) + \log
    \left|\det J_{F^{-1}} (\bx_a;\btheta,\bc_a)\right| \right\} \mmp{,}
  \end{aligned}
\end{equation}
where $\mathbb{E}_{\lhp[X](\bx,\bc)}$ denotes the expectation over the true joint distribution from which the training sample is drawn (approximated empirically by the second equality), and $F(\bx; \btheta, \bc)$ denotes the full network, parameterized by weights \btheta and conditioned on labels \bc. (The weights $\btheta$ should not be confused with the helicity angle $\theta$ introduced in the definition of the square Dalitz plot, cf. \cref{eq:m':theta'} in the main text.) For a two-dimensional Gaussian latent space
\begin{equation}
\mathcal{P}_{Z}(\bz) = (2\pi)^{-1} \exp \left( -\frac{1}{2}\, ||\bz||^{2}_{2} \right)
\end{equation}
we have
\begin{equation}
    - \log \lhp[Z]
    \left(F^{-1}(\bx_a; \btheta, \bc_a)\right) = \frac{1}{2} \bigl|\bigl|
  F^{-1}(\bx_a; \btheta, \bc_a)
  \bigr|\bigr|^2_2 + \log(2\pi).
\end{equation}
Dropping the $\mathbf{\theta}$-independent constant $\log(2\pi)$ from the loss  gives 
\begin{equation}
  \llh =  \frac{1}{N} \sum_{a=1}^N \left\{ \frac{1}{2}\bigl|\bigl|
  F^{-1}(\bx_a; \btheta, \bc_a)
  \bigr|\bigr|^2_2 - \log \bigl| \det J_F
  \left(F^{-1}(\bx_a;\btheta,\bc_a)\right)\bigr| \right\} \mmp{.}
  \label{eq:loss}
\end{equation}
where $|| \cdots ||^2_2$ denotes the squared $\ell^2$ norm and we have used the inverse function identity $\log |\det J_{F^{-1}}(\bx)|= - \log |\det J_F(\bz)|$. Because each step is differentiable, the gradient of \llh with respect to \btheta can be computed with standard automatic-differentiation tools.

In practice, the latent density is chosen so that it is easy to evaluate and sample, while the transformations $f_i$ must be sufficiently expressive and have tractable Jacobians. In our two-dimensional application we take the latent space variable to be normally distributed, i.e., $\bz \sim \mathcal{N}(\vec{0},1_{2\times2})$, where $\mathcal{N}$ is a two-dimensional Gaussian and $\vec{0}$ a two-dimensional zero vector. For the  maps $f_i$ we use rational-quadratic neural spline couplings~\cite{Durkan:2019nsq} as implemented in the \textsc{nflows} library~\cite{nflows}.

\subsection{Rational-quadratic spline coupling flows}

The rational-quadratic neural spline coupling construction of $f_i$ follows and extends the previous RealNVP construction~\cite{dinh2017density}, where at $i-$th step each coupling block splits the input into two channels. 
$\bz_{i-1} = \{\bz_{i-1,1}, \bz_{i-1,2}\}$ and applies a sequential monotonic transformation to each channel as follows
\begin{equation}
  \begin{aligned}
    &\bz_{i,1} = g_{i,1}\bigl(\bz_{i-1,1};\,\boldsymbol{\phi}_{i,1}(\bz_{i-1,2})\bigr) \mmp{,}\\
    &\bz_{i,2} = g_{i,2}\bigl(\bz_{i-1,2};\,\boldsymbol{\phi}_{i,2}(\bz_{i,1})\bigr) \mmp{.}
  \end{aligned}
  \label{eq:spline_transform}
\end{equation}
In our two-dimensional application, both $\bz_1$ and $\bz_2$ are one-dimensional. The assignment alternates between coupling blocks, and a random permutation of the components is applied after each block.
Here, each $g_{i,a}(\cdot\,;\boldsymbol{\phi})$ is a monotonic, piecewise rational-quadratic spline~\cite{Durkan:2019nsq,Gregory1982} whose spline parameters $\boldsymbol{\phi}$ (bin widths, bin heights, and knot derivatives) are predicted by a conditioner sub-network that takes the other channel as the input. The output $\bz_{i} = \{\bz_{i,1}, \bz_{i,2}\}$ is then passed to the next coupling block. Although not used in the present work, conditional information can be incorporated by concatenating labels $\bc$ to the inputs of the conditioner sub-networks, $\boldsymbol{\phi}_{i,a}(\bz) \to \boldsymbol{\phi}_{i,a}(\bz, \bc)$.

The spline $g_{i,a}(\cdot\,;\boldsymbol{\phi})$  partitions the real line into $N_b$ bins, defining $N_b{+}1$ knot points through which the transformation must pass. The conditioner sub-network predicts the widths and heights of these bins along with the derivative at each knot. Within each bin, the transformation is the unique rational-quadratic function (a ratio of two quadratic functions) that interpolates the positions and slopes at both bin boundaries~\cite{Gregory1982}. This is the simplest smooth monotonic interpolant that is analytically invertible, while requiring only the solution of a quadratic equation for the inverse map. Outside the outermost knots the transformation reduces to the identity.
The above piecewise construction is particularly well suited to describe Dalitz plot densities in $D$ decays, where narrow resonances produce rapid local variations that a single global transformation would struggle to capture. The spline instead can allocate knots densely in regions of rapid variation while remaining coarse elsewhere.

At initialization, the conditioner sub-network weights and biases are set to zero so that its output $\boldsymbol{\phi}$ vanishes. The spline parameters are then constructed by applying to appropriate components of $\boldsymbol{\phi}$ the softmax (for bin widths and heights) and softplus (for knot derivatives) functions. When $\boldsymbol{\phi}=\mathbf{0}$ the softmax produces uniform bins and the softplus yields unit derivatives, so that $g$ starts as the identity map. As training progresses the sub-network learns to produce nonzero $\boldsymbol{\phi}$, allowing the knot positions and slopes to deform the map away from the identity and capture the target density. The sub-network outputs are 
scaled by a factor of $0.30$ before constructing the spline parameters, which limits the initial deformation and stabilizes early training.

Each conditioner sub-network is a fully connected network with two hidden layers of 128 units each and SiLU activations. Its output is a vector of dimension $3N_b{-}1$, encoding the $N_b$ bin widths, $N_b$ bin heights, and $N_b{-}1$ interior knot derivatives. In our implementation $N_b=24$ and the spline tails extend beyond $\pm 5$.

 Because each $g_{i,a}$ in \cref{eq:spline_transform} is a monotonic spline, it is analytically invertible segment by segment. The inverse map is given by
\begin{equation}
  \begin{aligned}
    &\bz_{i-1,2} = g_{i,2}^{-1}\bigl(\bz_{i,2};\,\boldsymbol{\phi}_{i,2}(\bz_{i,1})\bigr) \mmp{,}\\
    &\bz_{i-1,1} = g_{i,1}^{-1}\bigl(\bz_{i,1};\,\boldsymbol{\phi}_{i,1}(\bz_{i-1,2})\bigr) \mmp{.}
  \end{aligned}
\end{equation}
By construction, the Jacobian matrix for each coupling block is triangular, so that the determinant of the full Jacobian $J_f$ is inexpensive to evaluate,
\begin{equation}
  \begin{aligned}
    \det J_f (\bz) = \det \frac{\partial f_{ij}}{\partial \bz}
    &= \det
    \begin{pmatrix}
      \text{diag} \left\{ g'_{i,1}\bigl(\bz_{i-1,1};\,\boldsymbol{\phi}_{i,1}\bigr) \right\}
      & \cdots \\
      0 & \text{diag} \left\{ g'_{i,2}\bigl(\bz_{i-1,2};\,\boldsymbol{\phi}_{i,2}\bigr) \right\}
    \end{pmatrix} \\
    &= \prod g'_{i,1}\bigl(\bz_{i-1,1};\,\boldsymbol{\phi}_{i,1}\bigr) \prod
    g'_{i,2}\bigl(\bz_{i-1,2};\,\boldsymbol{\phi}_{i,2}\bigr) \mmp{,}
  \end{aligned}
  \label{eq:det_jac_spline}
\end{equation}
where $g'_{i,a}$ denotes the derivative of the rational-quadratic spline with respect to its first argument, which is available analytically.

For the application to the $D\to K_S\pi^+\pi^-$ square Dalitz plot, before the first coupling block, the input coordinates $(m',\theta') \in [0,1]^2$ are mapped to $\mathbb{R}^2$ via the logit (inverse-sigmoid) transform. Our flow stacks 12 coupling blocks, each followed by a random permutation of the two components. Because each coupling layer leaves one channel unchanged, permuting the components between layers ensures that both coordinates are transformed in varied orderings across the stack, preventing the composition from developing blind spots in the input space. 

\subsection{Model hyperparameters and training}
\label{app:model_details}

Each of the three normalizing flows (flavor, CP-even, CP-odd) uses the same rational-quadratic neural spline flow (RQ-NSF)~\cite{Durkan:2019nsq} architecture, as implemented in the \textsc{nflows} library~\cite{nflows}. The flow consists of 12 coupling blocks with alternating masks and random permutations, each using a 2-layer conditioner sub-network with 128 hidden units, SiLU activations, and an output scale of $0.30$. The splines use $N_b = 24$ bins per dimension with linear tails beyond $\pm 5$, giving approximately $3.1\times10^5$ trainable parameters per flow. All flows are trained in square-Dalitz  coordinates $[0,1]^2$ with a logit boundary transform applied to the inputs.

Training uses the Adam optimizer with an initial learning rate of $9\times10^{-3}$.
The learning rate is reduced by a factor of 0.5 whenever the validation loss plateaus for 4 consecutive epochs (ReduceLROnPlateau scheduler), and training is stopped early if no improvement is seen for 30 epochs, up to a maximum of 450 epochs.
The batch size is 50{,}000 events and each dataset is split 90/10 into training and validation subsets.

\section{Constraint-aware extraction: training and architecture}
\label{app:constraint_aware}

This appendix provides the training details and architecture specification for the constraint-aware $h$-network introduced in \cref{sec:unbinned_fit}. The key idea and its motivation are described in the main text.

\subsection{Loss function and normalization}
\label{sec:loss:function:norm}

The network $h(m',\theta';\,\boldsymbol{\theta}_h)$ is trained by minimizing the joint negative log-likelihood on the CP-tagged samples. The normalized CP-tagged densities $p_\pm$ defined in \cref{eq:p+-:def} take the form 
\begin{align}
    p_\pm(m',\theta') &= \frac{1}{\hat{\Gamma}_\pm}\Big[K + \KK \pm 2\sqrt{K\KK}\;h\Big]\,,  \label{eq:phat_pm}
\end{align}
where $K$ and $\KK$ are the square-Dalitz  probability densities defined in \cref{eq:K_def}, and $\hat \Gamma_\pm$ the reduced decay widths for CP-even (CP-odd) tagged decays, cf. \cref{eq:Gammapm:h}. In the training of $h$ the $K$ function is frozen to its flavor flow value $K_\text{NF}$.  

Note that unlike a normalizing flow such as ${\mathcal C}$ in \cref{eq:C_extraction}, $p_\pm$ is not automatically normalized. The normalization constants $\Gamma_\pm$ depend on $h$ and must be tracked during training. Since $K$ is already a normalized density on the square-Dalitz plot ($\int K\,dm'd\theta' = 1$, and likewise for $\KK$), the structure simplifies: 
\begin{equation}
    \hat{\Gamma}_\pm = \bigg[\underbrace{\int (K + \KK)\,dm'\,d\theta'}_{=\;2\;\text{(fixed)}} \;\pm\; 2\underbrace{\int \sqrt{K\KK}\;h\;dm'\,d\theta'}_{\equiv\; I_h}\bigg] = 2(1 \pm I_h)\,.
\end{equation}
Only the single scalar $I_h$ must be recomputed as $h$ updates. It is estimated via a fixed, uniformly sampled Monte Carlo grid $\{(m'_\ell,\theta'_\ell)\}_{\ell=1}^{N_{\mathrm{MC}}}$ over $[0,1]^2$,
\begin{equation}\label{eq:I_h_MC}
    I_h \approx \frac{1}{N_{\mathrm{MC}}}\sum_{\ell=1}^{N_{\mathrm{MC}}} \sqrt{K_\ell\,\KK_\ell}\;h(m'_\ell,\theta'_\ell)\,.
\end{equation}
Since the weights $\sqrt{K_\ell\,\KK_\ell}$ depend only on the frozen flavor flow, they are fixed throughout training and can be precomputed once. Evaluating $I_h$ at each gradient step then reduces to a weighted sum over the current network outputs. 

The loss function is 
\begin{equation}\label{eq:h_loss}
    \mathcal{L}(\boldsymbol{\theta}_h) = -\sum_{j\in\mathcal{D}_+}\!\ln p_{+,j} \;-\; \sum_{k\in\mathcal{D}_-}\!\ln p_{-,k}\,,
\end{equation}
where $\mathcal{D}_\pm$ are the corresponding CP-tagged training samples. Training uses the Adam optimizer with a learning rate of $10^{-4}$ (reduced on plateau), a batch size of $20{,}000$, and early stopping with a patience of 30 epochs. The normalization $I_h$ is recomputed at each gradient step over the full MC grid ($N_{\mathrm{MC}} = 2.5\times10^5$) to avoid noisy normalization estimates.

After training, the interference observables entering the $B^\pm$ fit are \begin{equation}
\CC_\text{NF} = \sqrt{K_\text{NF} \KK_\text{NF}}\;h_\text{MLP}, \qquad |\SSS_\text{NF}| = \sqrt{K_\text{NF}\KK_\text{NF}}\;\sqrt{1-h_\text{MLP}^2},
\end{equation}
which satisfy $\CC_\text{NF}^2 + \SSS_\text{NF}^2 = K_\text{NF}\KK_\text{NF}$ for any $\boldsymbol{\theta}_h$. The sign of $\SSS_\text{NF}$ is determined from the amplitude model.

\subsection{\texorpdfstring{Architecture of $h$}{Architecture of h}}
\label{app:h}
For $h$ 
we use a
multilayer perceptron (MLP) mapping $[0,1]^2 \to [-1,1]$, with a $\tanh$ output activation that guarantees $|h|\leq 1$.
The challenge here is that the target function $\cos\Dd$ varies rapidly near narrow resonance crossings, e.g.,\ $K^*(892)$ in invariant $K_S\pi$ mass and $\rho(770)$ in invariant $\pi^+\pi^-$ mass distribution, creating sharp features in the Dalitz plot. Standard activation functions such as ReLU or SiLU exhibit a well-known spectral bias~\cite{Rahaman2019spectral}; they learn low-frequency components of the target function first and converge slowly, if at all, to high-frequency features. For Dalitz-plot applications this is problematic, since the interference pattern changes the sign over intervals comparable to the resonance width, which should be correctly captured by the MLP.

\begin{figure*}[t!]
    \centering
    \includegraphics[width=\textwidth]{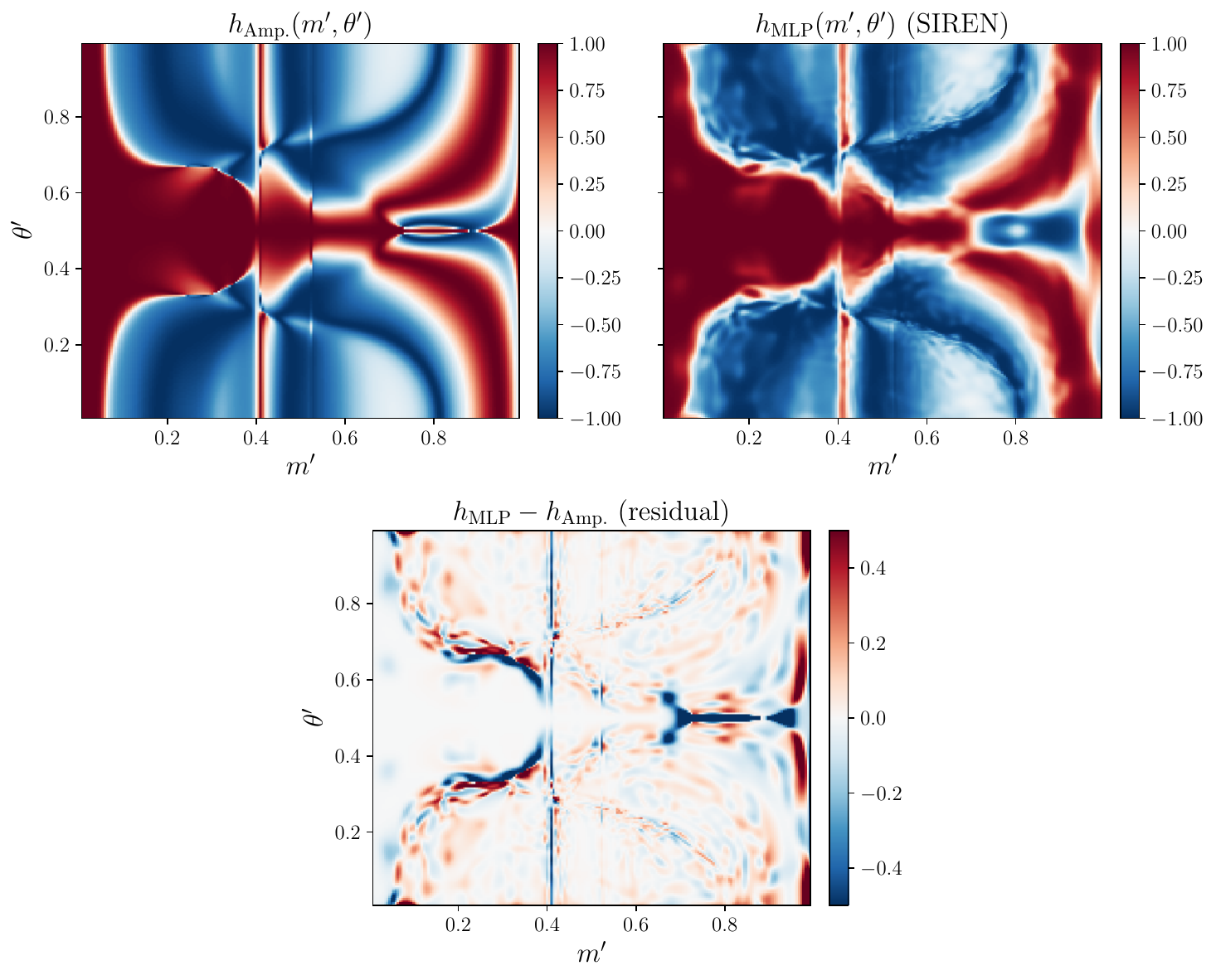}
    \caption{Comparison across the square Dalitz plot of $\cos\Dd$  from the amplitude model ($h_{\text{Amp.}}$, upper left) with the trained MLP prediction ($h_{\text{MLP}}$, upper right). The bottom panel shows the residual $h_{\text{MLP}} - h_{\text{Amp.}}$. The largest deviations are localized in regions of narrow resonance crossings.
    }
    \label{fig:h_vs_truth}
\end{figure*}

We therefore adopt a SIREN (Sinusoidal Representation Network)~\cite{Sitzmann2020siren} architecture, which replaces the standard activation with a sinusoid. That is, each hidden layer applies a transformation 
\begin{equation}
    \bz_{i} = \sin\,\bigl(\omega_0\,(\mathbf{W}_i\,\bz_{i-1} + \mathbf{b}_i)\bigr)\,,
\end{equation}
to the inputs $\bz_{i-1}$ from the previous layer. 
The frequency hyperparameter $\omega_0$ controls the characteristic oscillation scale. SIREN learns which frequencies are needed through its weights, and the multiplicative composition of $\sin$ across layers produces rich frequency combinations that can resolve sharp structures where they occur, while remaining smooth elsewhere.

\begin{figure*}[t]
    \centering
    \includegraphics[width=\textwidth]{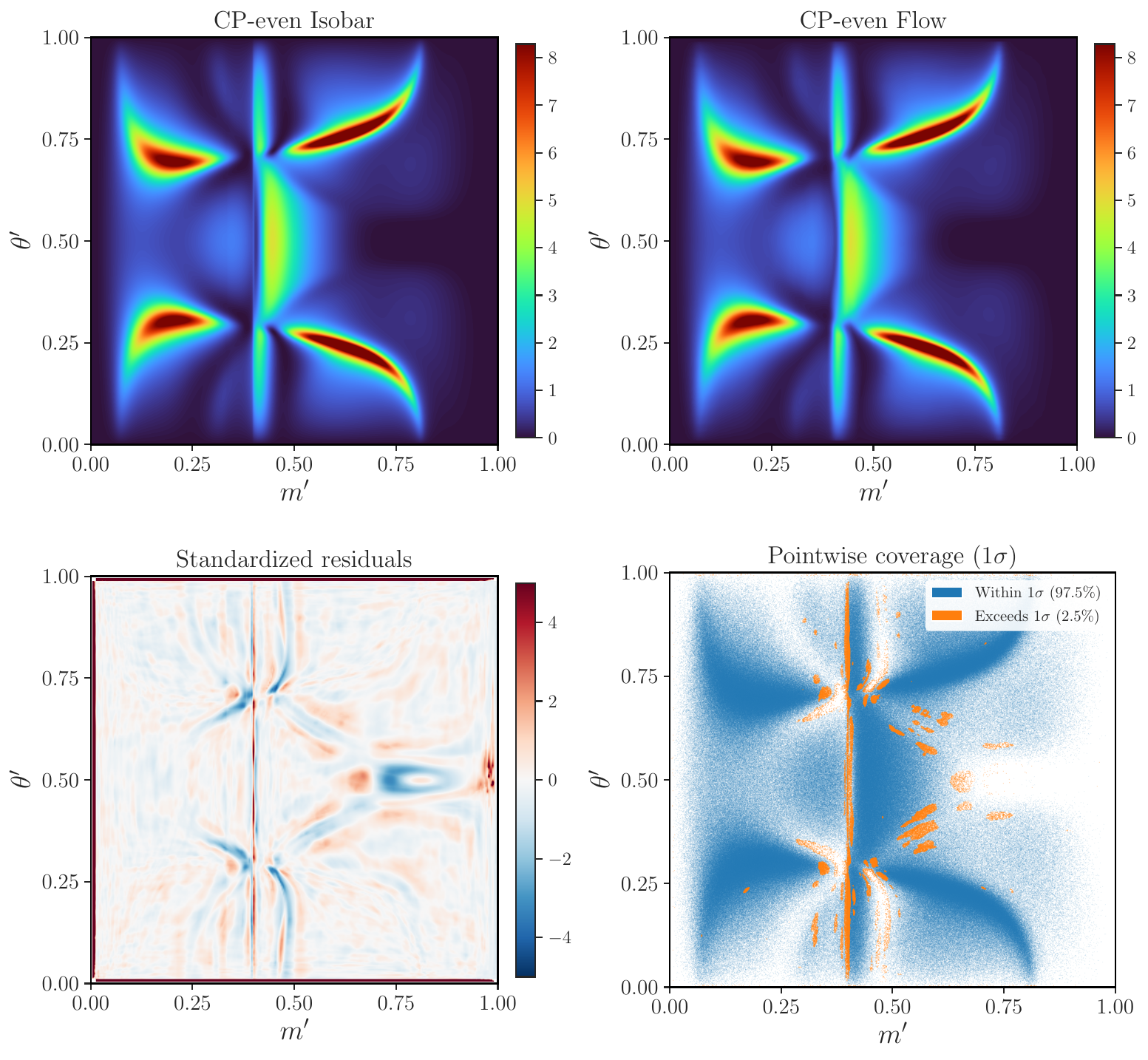}
    
    \caption{Coverage diagnostics for the CP-even flow ensemble of $N_\text{ens}^D=25$ independently trained models. {\em Upper left:}~Isobar-model density $\rho_{\rm iso}(m',\theta')$. {\em Upper right:}~Flow ensemble mean density $\langle\rho(m',\theta')\rangle$. {\em Bottom left:}~Standardized residuals $(\langle\rho\rangle - \rho_{\rm iso})/\sigma_{\rm ens}$; deviations beyond $\pm 2$ are concentrated along narrow resonance bands. {\em Bottom right:}~Pointwise $1\sigma$ coverage map: within $1\sigma$ (blue, $97.5\%$), and exceeding $1\sigma$ (orange, $2.5\%$).}
    \label{fig:coverage_test_even}
\end{figure*}

The SIREN weights are initialized following the scheme of Ref.~\cite{Sitzmann2020siren}. The first-layer weights are drawn uniformly from the interval $[-1/n,\, 1/n]$, while the subsequent layers from $[-\sqrt{6/n}/\omega_0,\, \sqrt{6/n}/\omega_0]$, where $n$ is the input dimension of each layer. This ensures that the pre-activations remain in the approximately linear regime of $\sin$ at initialization.

In the symmetric square Dalitz plot with the $\pi^+\pi^-$ resonant-pair convention, the function $\cos\Dd$ is symmetric under $\theta' \to 1-\theta'$ (since swapping $\pi^+ \leftrightarrow \pi^-$ leaves $\mathrm{Re}(A_D\bar{A}_D^*)$ invariant). In the NN architecture we enforce this symmetry exactly by averaging the raw network output,
\begin{equation}\label{eq:h_sym}
    h(m',\theta') = \tfrac{1}{2}\bigl[g(m',\theta') + g(m',1{-}\theta')\bigr]\,,
\end{equation}
where $g$ is the underlying SIREN. This halves the effective function space that must be learned and eliminates any antisymmetric artifacts. The specific architecture used in this work consists of 5 hidden layers of 256 units each with $\omega_0 = 15$, giving approximately $2\times 10^5$ trainable parameters.

\begin{figure*}[t!]
    \centering
    \includegraphics[width=\textwidth]{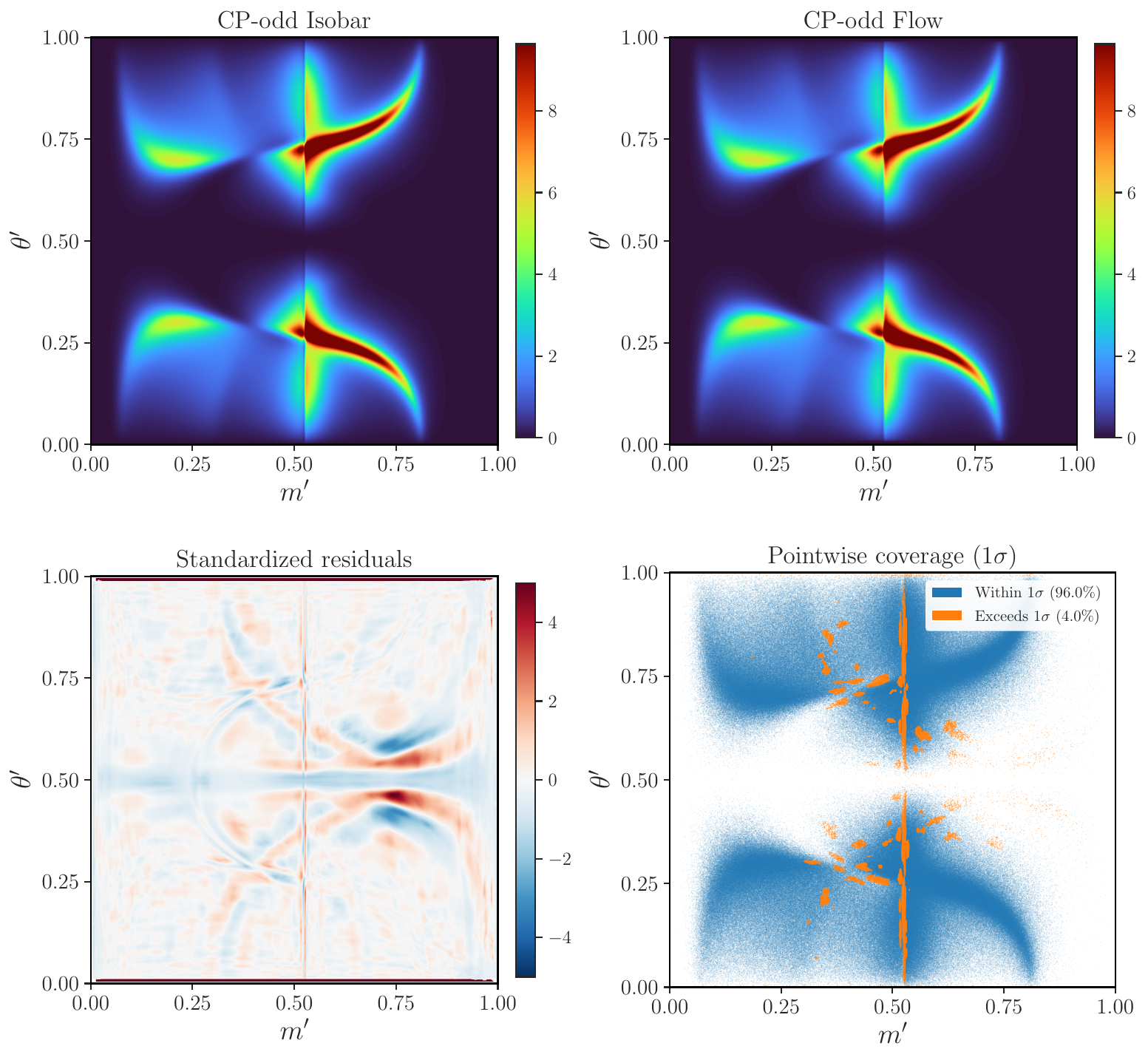}
    \caption{Coverage diagnostics for the CP-odd flow ensemble of $N_\text{ens}^D=25$ independently trained models. {\em Upper left:}~Isobar-model density $\rho_{\rm iso}(m',\theta')$. {\em Upper right:}~Flow ensemble mean density $\langle\rho(m',\theta')\rangle$. {\em Bottom left:}~Standardized residuals $(\langle\rho\rangle - \rho_{\rm iso})/\sigma_{\rm ens}$; deviations beyond $\pm 2$ are concentrated along narrow resonance bands. {\em Bottom right:}~Pointwise $1\sigma$ coverage map: within $1\sigma$ (blue, $96.0\%$), and exceeding $1\sigma$ (orange, $4.0\%$).}
    \label{fig:coverage_test_odd}
\end{figure*}

During development, several alternative architectures were explored. While none were exhaustively optimized, the patterns observed may be useful for future implementations. MLPs with Fourier positional encodings~\cite{Tancik2020fourier} appended to the input were found to be unstable under the indirect NLL loss of \cref{eq:h_loss}, because high-frequency basis functions can push the model densities $\hat{p}_\pm$ negative, causing numerical divergences. Residual networks with zero-initialized output blocks stabilized training but offered no improvement over plain MLPs. A two-flow density-ratio approach, $h = \tanh[(\log q_1 - \log q_2)/2]$, achieved competitive NLL values but produced noisy, spiky estimates of $h$ due to amplification of statistical fluctuations in the ratio. Adding gradient penalties to smooth the ratio was ineffective because the piecewise-polynomial structure of the spline transforms creates large gradient variance at bin boundaries. Among the architectures tested, SIREN performed best. Its sinusoidal activations provide the necessary frequency content without external encodings, and its smooth, weight-sharing structure provides implicit regularization against overfitting.

\Cref{fig:h_vs_truth} compares the output of a trained $h$-network with the one from the amplitude model across the square Dalitz plot. The network reproduces the large-scale structure faithfully, with residuals along narrow resonance crossings where $\cos\Dd$ changes rapidly.

\section{CP-tagged flow coverage}
\label{app:cp_coverage}

This appendix presents the pointwise coverage diagnostics for the CP-even and CP-odd flows $p_{\pm}$, complementing the 3-flow analysis in \cref{sec:coverage}.

\Cref{fig:coverage_test_even} shows the results for the CP-even flow. The ensemble mean density reproduces the isobar-model resonance structure across the full square Dalitz plot. The $1\sigma$ coverage fraction is $97.5\%$, with the remaining $2.5\%$ of grid points lying outside the expected $1\sigma$ error band. Standardized residuals exceeding $\pm 2$ are localized to narrow resonance bands, consistent with the flavor-flow behavior.

\Cref{fig:coverage_test_odd} shows the corresponding results for the CP-odd flow. The $1\sigma$ coverage fraction is $96.0\%$, with the remaining $4.0\%$ of grid points lying outside the expected $1\sigma$ error band. The pattern of coverage failures is again concentrated along narrow resonance structures.

\end{appendix}

\bibliographystyle{JHEP}
\bibliography{GMSSZ}

@article{Gronau:2001nr,
    author = "Gronau, Michael and Grossman, Yuval and Rosner, Jonathan L.",
    title = "{Measuring D0 - anti-D0 mixing and relative strong phases at a charm factory}",
    eprint = "hep-ph/0103110",
    archivePrefix = "arXiv",
    reportNumber = "TECHNION-PH-2001-19, EFI-01-07",
    doi = "10.1016/S0370-2693(01)00426-9",
    journal = "Phys. Lett. B",
    volume = "508",
    pages = "37--43",
    year = "2001"
}

@article{James:1975dr,
    author = "James, F. and Roos, M.",
    title = "{Minuit: A System for Function Minimization and Analysis of the Parameter Errors and Correlations}",
    reportNumber = "CERN-DD-75-20",
    doi = "10.1016/0010-4655(75)90039-9",
    journal = "Comput. Phys. Commun.",
    volume = "10",
    pages = "343--367",
    year = "1975"
}

@article{Aston:1987ir,
    author = "Aston, D. and others",
    title = "{A Study of K- pi+ Scattering in the Reaction K- p ---{\ensuremath{>}} K- pi+ n at 11-GeV/c}",
    reportNumber = "SLAC-PUB-4260, DPNU-87-25",
    doi = "10.1016/0550-3213(88)90028-4",
    journal = "Nucl. Phys. B",
    volume = "296",
    pages = "493--526",
    year = "1988"
}

@article{Ernst:2023qvn,
    author = "Ernst, Florian and Favaro, Luigi and Krause, Claudius and Plehn, Tilman and Shih, David",
    title = "{Normalizing flows for high-dimensional detector simulations}",
    eprint = "2312.09290",
    archivePrefix = "arXiv",
    primaryClass = "hep-ph",
    doi = "10.21468/SciPostPhys.18.3.081",
    journal = "SciPost Phys.",
    volume = "18",
    number = "3",
    pages = "081",
    year = "2025"
}

@article{Bierlich:2023zzd,
    author = "Bierlich, Christian and Ilten, Phil and Menzo, Tony and Mrenna, Stephen and Szewc, Manuel and Wilkinson, Michael K. and Youssef, Ahmed and Zupan, Jure",
    title = "{Towards a data-driven model of hadronization using normalizing flows}",
    eprint = "2311.09296",
    archivePrefix = "arXiv",
    primaryClass = "hep-ph",
    reportNumber = "FERMILAB-PUB-23-698-CSAID",
    doi = "10.21468/SciPostPhys.17.2.045",
    journal = "SciPost Phys.",
    volume = "17",
    number = "2",
    pages = "045",
    year = "2024"
}

@article{Butter:2025wxn,
    author = "Butter, Anja and others",
    title = "{Iterative HOMER with uncertainties}",
    eprint = "2509.03592",
    archivePrefix = "arXiv",
    primaryClass = "hep-ph",
    reportNumber = "FERMILAB-PUB-25-0579-CSAID",
    doi = "10.21468/SciPostPhys.20.2.042",
    journal = "SciPost Phys.",
    volume = "20",
    number = "2",
    pages = "042",
    year = "2026"
}

@phdthesis{neal1995bayesian,
  author  = {Neal, Radford M.},
  title   = {Bayesian Learning for Neural Networks},
  school  = {University of Toronto},
  year    = {1995},
  url     = {ftp://www.cs.toronto.edu/dist/radford/thesis.pdf}
}

@phdthesis{gal2016uncertainty,
  author  = {Gal, Yarin},
  title   = {Uncertainty in Deep Learning},
  school  = {University of Cambridge},
  year    = {2016},
  url     = {http://mlg.eng.cam.ac.uk/yarin/thesis/thesis.pdf}
}

@article{Bollweg:2019skg,
    author = "Bollweg, Sven and Hau{\ss}mann, Manuel and Kasieczka, Gregor and Luchmann, Michel and Plehn, Tilman and Thompson, Jennifer",
    title = "{Deep-Learning Jets with Uncertainties and More}",
    eprint = "1904.10004",
    archivePrefix = "arXiv",
    primaryClass = "hep-ph",
    doi = "10.21468/SciPostPhys.8.1.006",
    journal = "SciPost Phys.",
    volume = "8",
    number = "1",
    pages = "006",
    year = "2020"
}

@article{mackay1995probable,
  author    = {MacKay, David J. C.},
  title     = {Probable Networks and Plausible Predictions -- A Review of Practical {Bayesian} Methods for Supervised Neural Networks},
  journal   = {Network: Computation in Neural Systems},
  volume    = {6},
  number    = {3},
  pages     = {469--505},
  year      = {1995},
  publisher = {Taylor \& Francis},
  doi       = {10.1088/0954-898X_6_3_011}
}

@article{HeavyFlavorAveragingGroupHFLAV:2024ctg,
    author = "Banerjee, Sw. and others",
    collaboration = "Heavy Flavor Averaging Group (HFLAV)",
    title = "{Averages of b-hadron, c-hadron, and {\ensuremath{\tau}}-lepton properties as of 2023}",
    eprint = "2411.18639",
    archivePrefix = "arXiv",
    primaryClass = "hep-ex",
    doi = "10.1103/x87q-tld5",
    journal = "Phys. Rev. D",
    volume = "113",
    number = "1",
    pages = "012008",
    year = "2026"
}

@article{Bondar:2008hh,
    author = "Bondar, A. and Poluektov, A.",
    title = "{The Use of quantum-correlated D0 decays for phi3 measurement}",
    eprint = "0801.0840",
    archivePrefix = "arXiv",
    primaryClass = "hep-ex",
    doi = "10.1140/epjc/s10052-008-0600-z",
    journal = "Eur. Phys. J. C",
    volume = "55",
    pages = "51--56",
    year = "2008"
}

@article{BaBar:2007hmp,
    author = "Aubert, Bernard and others",
    collaboration = "BaBar",
    title = "{Dalitz Plot Analysis of the Decay $B^0$ (anti-B0) $\to K^\pm \pi^\mp \pi^0$}",
    eprint = "0711.4417",
    archivePrefix = "arXiv",
    primaryClass = "hep-ex",
    reportNumber = "BABAR-PUB-07-066, SLAC-PUB-13023",
    doi = "10.1103/PhysRevD.78.052005",
    journal = "Phys. Rev. D",
    volume = "78",
    pages = "052005",
    year = "2008"
}

@article{Back:2017zqt,
    author = "Back, John and others",
    title = "{LAURA$^{++}$: A Dalitz plot fitter}",
    eprint = "1711.09854",
    archivePrefix = "arXiv",
    primaryClass = "hep-ex",
    doi = "10.1016/j.cpc.2018.04.017",
    journal = "Comput. Phys. Commun.",
    volume = "231",
    pages = "198--242",
    year = "2018"
}

@misc{HFLAV:GammaSummer2025,
  author       = {{HFLAV Collaboration}},
  title        = {{Results on Time-Dependent CP Violation and Measurements Related to the Angles of the Unitarity Triangle}},
  howpublished = {\url{https://hflav-eos.web.cern.ch/hflav-eos/triangle/summer2025/\#gamma_comb}},
  year         = {2025},
  note         = {Summer 2025 averages and combinations (EPS 2025, Lepton Photon 2025, CKM2025, etc.)},
}

@inproceedings{Grossman:2017thq,
    author = "Grossman, Yuval and Tanedo, Philip",
    title = "{Just a taste: lectures on flavor physics.}",
    booktitle = "{Theoretical Advanced Study Institute in Elementary Particle Physics}: {Anticipating the Next Discoveries in Particle Physics}",
    eprint = "1711.03624",
    archivePrefix = "arXiv",
    primaryClass = "hep-ph",
    reportNumber = "UCR-TR-2017-FLIP-K-2SO",
    doi = "10.1142/9789813233348_0004",
    pages = "109--295",
    year = "2018"
}

@article{Lane:2023iak,
    author = "Lane, Jake and Gersabeck, Evelina and Rademacker, Jonas",
    title = "{A novel unbinned model-independent method to measure the CKM angle {\ensuremath{\gamma}} in B$^{±}$ {\textrightarrow} DK$^{±}$ decays with optimised precision}",
    eprint = "2305.10787",
    archivePrefix = "arXiv",
    primaryClass = "hep-ph",
    doi = "10.1007/JHEP09(2023)007",
    journal = "JHEP",
    volume = "09",
    pages = "007",
    year = "2023"
}

@article{Brod:2014qwa,
    author = "Brod, Joachim",
    title = "{Electroweak effects in the extraction of the CKM angle $\gamma$ from B$\to D \pi$ decays}",
    eprint = "1412.3173",
    archivePrefix = "arXiv",
    primaryClass = "hep-ph",
    reportNumber = "MITP-14-098",
    doi = "10.1016/j.physletb.2015.02.022",
    journal = "Phys. Lett. B",
    volume = "743",
    pages = "56--60",
    year = "2015"
}

@article{Zupan:2019uoi,
    author = "Zupan, Jure",
    editor = "Mulders, M. and Duhr, C.",
    title = "{Introduction to flavour physics}",
    eprint = "1903.05062",
    archivePrefix = "arXiv",
    primaryClass = "hep-ph",
    doi = "10.23730/CYRSP-2019-006.181",
    journal = "CERN Yellow Rep. School Proc.",
    volume = "6",
    pages = "181--212",
    year = "2019"
}

@article{BESIII:2020khq,
    author = "Ablikim, M. and others",
    collaboration = "BESIII",
    title = "{Model-independent determination of the relative strong-phase difference between $D^0$ and $\bar{D}^0\rightarrow K^0_{S,L}\pi^+\pi^-$ and its impact on the measurement of the CKM angle $\gamma/\phi_3$}",
    eprint = "2003.00091",
    archivePrefix = "arXiv",
    primaryClass = "hep-ex",
    doi = "10.1103/PhysRevD.101.112002",
    journal = "Phys. Rev. D",
    volume = "101",
    number = "11",
    pages = "112002",
    year = "2020"
}

@inproceedings{Bondar:2007ir,
    author = "Bondar, Alex and Poluektov, Anton",
    title = "{On model-independent measurement of the angle phi(3) using Dalitz plot analysis}",
    booktitle = "{4th International Workshop on the CKM Unitarity Triangle (CKM 2006)}",
    eprint = "hep-ph/0703267",
    archivePrefix = "arXiv",
    month = "3",
    year = "2007"
}

@article{Bondar:2005ki,
    author = "Bondar, A. and Poluektov, A.",
    title = "{Feasibility study of model-independent approach to phi(3) measurement using Dalitz plot analysis}",
    eprint = "hep-ph/0510246",
    archivePrefix = "arXiv",
    doi = "10.1140/epjc/s2006-02590-x",
    journal = "Eur. Phys. J. C",
    volume = "47",
    pages = "347--353",
    year = "2006"
}

@article{Brod:2013sga,
    author = "Brod, Joachim and Zupan, Jure",
    title = "{The ultimate theoretical error on $\gamma$ from $B \to DK$ decays}",
    eprint = "1308.5663",
    archivePrefix = "arXiv",
    primaryClass = "hep-ph",
    doi = "10.1007/JHEP01(2014)051",
    journal = "JHEP",
    volume = "01",
    pages = "051",
    year = "2014"
}

@article{LHCb:2021dcr,
    author = "Aaij, Roel and others",
    collaboration = "LHCb",
    title = "{Simultaneous determination of CKM angle $\gamma$ and charm mixing parameters}",
    eprint = "2110.02350",
    archivePrefix = "arXiv",
    primaryClass = "hep-ex",
    reportNumber = "LHCb-PAPER-2021-033, CERN-EP-2021-183",
    doi = "10.1007/JHEP12(2021)141",
    journal = "JHEP",
    volume = "12",
    pages = "141",
    year = "2021"
}

@article{Belle:2004bbr,
    author = "Poluektov, A. and others",
    collaboration = "Belle",
    title = "{Measurement of phi(3) with Dalitz plot analysis of B+- ---\ensuremath{>} D**(*) K+- decay}",
    eprint = "hep-ex/0406067",
    archivePrefix = "arXiv",
    doi = "10.1103/PhysRevD.70.072003",
    journal = "Phys. Rev. D",
    volume = "70",
    pages = "072003",
    year = "2004"
}

@article{LHCb:2020yot,
    author = "Aaij, R. and others",
    collaboration = "LHCb",
    title = "{Measurement of the CKM angle $\gamma$ in $B^\pm\to D K^\pm$ and $B^\pm \to D \pi^\pm$ decays with $D \to K_\mathrm S^0 h^+ h^-$}",
    eprint = "2010.08483",
    archivePrefix = "arXiv",
    primaryClass = "hep-ex",
    reportNumber = "LHCb-PAPER-2020-019, CERN-EP-2020-175",
    doi = "10.1007/JHEP02(2021)169",
    journal = "JHEP",
    volume = "02",
    pages = "169",
    year = "2021"
}

@article{Gronau:1991dp,
    author = "Gronau, Michael and Wyler, Daniel",
    title = "{On determining a weak phase from CP asymmetries in charged B decays}",
    reportNumber = "TECHNION-PH-91-14, ZU-TH-6-91",
    doi = "10.1016/0370-2693(91)90034-N",
    journal = "Phys. Lett. B",
    volume = "265",
    pages = "172--176",
    year = "1991"
}

@inproceedings{Bondar:2002,
    author = "Bondar, Alex",
    title = "{Improved Gronau--Wyler method for $\phi_3$ extraction}",
    booktitle = "BINP special analysis meeting on Dalitz analysis, Sep. 24--26",
    month = "9",
    year = "2002",
    note = "unpublished"
}

@article{Ceccucci:2020cim,
    author = "Ceccucci, A. and Gershon, T. and Kenzie, M. and Ligeti, Z. and Sakai, Y. and Trabelsi, K.",
    title = "{Origins of the method to determine the CKM angle $\gamma$ using $B^{\pm} \to D K^{\pm}$, $D \to K_{\rm S}^0\pi^+\pi^-$ decays}",
    eprint = "2006.12404",
    archivePrefix = "arXiv",
    primaryClass = "physics.hist-ph",
    month = "6",
    year = "2020"
}

@article{BaBar:2018cka,
    author = "Adachi, I. and others",
    collaboration = "BaBar, Belle",
    title = "{Measurement of $\cos{2\beta}$ in $B^{0} \to D^{(*)} h^{0}$ with $D \to K_{S}^{0} \pi^{+} \pi^{-}$ decays by a combined time-dependent Dalitz plot analysis of BaBar and Belle data}",
    eprint = "1804.06153",
    archivePrefix = "arXiv",
    primaryClass = "hep-ex",
    doi = "10.1103/PhysRevD.98.112012",
    journal = "Phys. Rev. D",
    volume = "98",
    number = "11",
    pages = "112012",
    year = "2018"
}

@article{Poluektov:2017zxp,
    author = "Poluektov, Anton",
    title = "{Unbinned model-independent measurements with coherent admixtures of multibody neutral $D$ meson decays}",
    eprint = "1712.08326",
    archivePrefix = "arXiv",
    primaryClass = "hep-ph",
    doi = "10.1140/epjc/s10052-018-5599-1",
    journal = "Eur. Phys. J. C",
    volume = "78",
    number = "2",
    pages = "121",
    year = "2018"
}

@article{Grossman:2005rp,
    author = "Grossman, Yuval and Soffer, Abner and Zupan, Jure",
    title = "{The Effect of $D-\bar{D}$ mixing on the measurement of $\gamma$ in $B\to DK$ decays}",
    eprint = "hep-ph/0505270",
    archivePrefix = "arXiv",
    doi = "10.1103/PhysRevD.72.031501",
    journal = "Phys. Rev. D",
    volume = "72",
    pages = "031501",
    year = "2005"
}

@article{Giri:2003ty,
    author = "Giri, Anjan and Grossman, Yuval and Soffer, Abner and Zupan, Jure",
    title = "{Determining gamma using $B^\pm \to DK^\pm$ with multibody $D$ decays}",
    eprint = "hep-ph/0303187",
    archivePrefix = "arXiv",
    doi = "10.1103/PhysRevD.68.054018",
    journal = "Phys.\ Rev.\ D",
    volume = "68",
    pages = "054018",
    year = "2003"
}

@article{Gronau:1990ra,
    author = "Gronau, Michael and London, David",
    title = "{How to determine all the angles of the unitarity triangle from $B_d^0 \to D K_s$ and $B_s^0 \to D_\phi$}",
    reportNumber = "DESY-90-125, TECHNION-PH-90-26, UDEM-LPN-TH-34",
    doi = "10.1016/0370-2693(91)91756-L",
    journal = "Phys. Lett. B",
    volume = "253",
    pages = "483--488",
    year = "1991"
}

@article{Carter:1980tk,
    author = "Carter, Ashton B. and Sanda, A. I.",
    title = "{CP Violation in B Meson Decays}",
    reportNumber = "DOE/EY/2232B-205",
    doi = "10.1103/PhysRevD.23.1567",
    journal = "Phys. Rev. D",
    volume = "23",
    pages = "1567",
    year = "1981"
}

@article{Carter:1980hr,
    author = "Carter, Ashton B. and Sanda, A. I.",
    title = "{CP Violation in Cascade Decays of B Mesons}",
    reportNumber = "DOE/EY/2232B-203",
    doi = "10.1103/PhysRevLett.45.952",
    journal = "Phys. Rev. Lett.",
    volume = "45",
    pages = "952",
    year = "1980"
}

@article{CLEO:2002uvu,
    author = "Muramatsu, H. and others",
    collaboration = "CLEO",
    title = "{Dalitz analysis of D0 ---{\ensuremath{>}} K0(S) pi+ pi-}",
    eprint = "hep-ex/0207067",
    archivePrefix = "arXiv",
    reportNumber = "CLNS-02-1792, CLEO-02-10",
    doi = "10.1103/PhysRevLett.89.251802",
    journal = "Phys. Rev. Lett.",
    volume = "89",
    pages = "251802",
    year = "2002",
    note = "[Erratum: Phys.Rev.Lett. 90, 059901 (2003)]"
}

@article{Belle:2007tti,
    author = "Abe, Kazuo and others",
    editor = "Frere, Jean Marie and Iconomidou-Fayard, Lydia and Montanet, Francois and Tran Thanh Van, Jean",
    collaboration = "Belle",
    title = "{Measurement of $D^0 - \bar {D^0}$ Mixing Parameters in $D^0 \to K_s^0 \pi^+ \pi^-$ decays}",
    eprint = "0704.1000",
    archivePrefix = "arXiv",
    primaryClass = "hep-ex",
    reportNumber = "BELLE-CONF-0702",
    doi = "10.1103/PhysRevLett.99.131803",
    journal = "Phys. Rev. Lett.",
    volume = "99",
    pages = "131803",
    year = "2007"
}

@article{BaBar:2010nhz,
    author = "del Amo Sanchez, P. and others",
    collaboration = "BaBar",
    title = "{Measurement of D0-antiD0 mixing parameters using D0 ---{\ensuremath{>}} K(S)0 pi+ pi- and D0 ---{\ensuremath{>}} K(S)0 K+ K- decays}",
    eprint = "1004.5053",
    archivePrefix = "arXiv",
    primaryClass = "hep-ex",
    reportNumber = "BABAR-PUB-10-004, SLAC-PUB-14087",
    doi = "10.1103/PhysRevLett.105.081803",
    journal = "Phys. Rev. Lett.",
    volume = "105",
    pages = "081803",
    year = "2010"
}

@inproceedings{Reichert:2013ewa,
    author = "Reichert, Stefanie",
    collaboration = "LHCb",
    title = "{Time-dependent amplitude analysis of semileptonically-tagged D0 -{\ensuremath{>}} KS0 pi+ pi- decays at LHCb}",
    booktitle = "{6th International Workshop on Charm Physics}",
    eprint = "1311.2449",
    archivePrefix = "arXiv",
    primaryClass = "hep-ex",
    month = "11",
    year = "2013"
}

@article{Belle:2014ydf,
    author = "Peng, T. and others",
    collaboration = "Belle",
    title = "{Measurement of $D^0-\bar{D}^0$ mixing and search for indirect CP violation using $D^0\to K_S^0\pi^+\pi^-$ decays}",
    eprint = "1404.2412",
    archivePrefix = "arXiv",
    primaryClass = "hep-ex",
    doi = "10.1103/PhysRevD.89.091103",
    journal = "Phys. Rev. D",
    volume = "89",
    number = "9",
    pages = "091103",
    year = "2014"
}

@article{CLEO:2010iul,
    author = "Libby, J. and others",
    collaboration = "CLEO",
    title = "{Model-independent determination of the strong-phase difference between $D^0$ and $\bar{D}^0 \to K^0_{S,L} h^+ h^-$ ($h=\pi,K$) and its impact on the measurement of the CKM angle $\gamma/\phi_3$}",
    eprint = "1010.2817",
    archivePrefix = "arXiv",
    primaryClass = "hep-ex",
    reportNumber = "CLNS-10-2070, CLEO-10-07",
    doi = "10.1103/PhysRevD.82.112006",
    journal = "Phys. Rev. D",
    volume = "82",
    pages = "112006",
    year = "2010"
}

@article{Kobyzev_2021,
   title={Normalizing Flows: An Introduction and Review of Current Methods},
   volume={43},
   ISSN={1939-3539},
   url={http://dx.doi.org/10.1109/TPAMI.2020.2992934},
   DOI={10.1109/tpami.2020.2992934},
   number={11},
   journal={IEEE Transactions on Pattern Analysis and Machine Intelligence},
   publisher={Institute of Electrical and Electronics Engineers (IEEE)},
   author={Kobyzev, Ivan and Prince, Simon J.D. and Brubaker, Marcus A.},
   year={2021},
   month=nov, pages={3964-3979} }

@misc{Rezende:2015a,
      title={Variational Inference with Normalizing Flows}, 
      author={Danilo Jimenez Rezende and Shakir Mohamed},
      year={2016},
      eprint={1505.05770},
      archivePrefix={arXiv},
      primaryClass={stat.ML},
      url={https://arxiv.org/abs/1505.05770}, 
}

@article{Backus:2022xhi,
    author = "Backus, Jeffrey V. and Freytsis, Marat and Grossman, Yuval and Schacht, Stefan and Zupan, Jure",
    title = "{Toward extracting $\gamma $ from $B\rightarrow DK$ without binning}",
    eprint = "2211.05133",
    archivePrefix = "arXiv",
    primaryClass = "hep-ph",
    doi = "10.1140/epjc/s10052-023-12057-x",
    journal = "Eur. Phys. J. C",
    volume = "83",
    pages = "877",
    year = "2023"
}

@misc{dinh2015,
      title={NICE: Non-linear Independent Components Estimation},
      author={Laurent Dinh and David Krueger and Yoshua Bengio},
      year={2015},
      eprint={1410.8516},
      archivePrefix={arXiv},
      primaryClass={cs.LG},
      url={https://arxiv.org/abs/1410.8516},
}

@inproceedings{dinh2017density,
    author    = "Dinh, Laurent and Sohl-Dickstein, Jascha and Bengio, Samy",
    title     = "{Density estimation using Real-NVP}",
    booktitle = "International Conference on Learning Representations",
    year      = "2017",
    eprint    = "1605.08803",
    archivePrefix = "arXiv",
    primaryClass  = "cs.LG"
}

@software{nflows,
    author    = "Durkan, Conor and Bekasov, Artur and Murray, Iain and
                 Papamakarios, George",
    title     = "{nflows: normalizing flows in PyTorch}",
    year      = "2020",
    doi       = "10.5281/zenodo.4296287",
    url       = "https://github.com/bayesiains/nflows"
}

@inproceedings{Durkan:2019nsq,
    author        = "Durkan, Conor and Bekasov, Artur and Murray, Iain and
                     Papamakarios, George",
    title         = "{Neural Spline Flows}",
    booktitle     = "Advances in Neural Information Processing Systems",
    volume        = "32",
    year          = "2019",
    publisher     = "Curran Associates, Inc.",
    eprint        = "1906.04032",
    archivePrefix = "arXiv",
    primaryClass  = "stat.ML"
}

@inproceedings{Louizos2017mnf,
    author    = "Louizos, Christos and Welling, Max",
    title     = "{Multiplicative Normalizing Flows for Variational Bayesian Neural Networks}",
    booktitle = "International Conference on Machine Learning",
    year      = "2017",
    eprint    = "1703.01961",
    archivePrefix = "arXiv",
    primaryClass  = "stat.ML"
}

@inproceedings{Tancik2020fourier,
    author    = "Tancik, Matthew and Srinivasan, Pratul P. and Mildenhall, Ben and
                 Fridovich-Keil, Sara and Raghavan, Nithin and Singhal, Utkarsh and
                 Ramamoorthi, Ravi and Barron, Jonathan T. and Ng, Ren",
    title     = "{Fourier Features Let Networks Learn High Frequency Functions in Low Dimensional Domains}",
    booktitle = "Advances in Neural Information Processing Systems",
    volume    = "33",
    year      = "2020",
    eprint    = "2006.10739",
    archivePrefix = "arXiv",
    primaryClass  = "cs.CV"
}

@inproceedings{Rahaman2019spectral,
    author    = "Rahaman, Nasim and Baratin, Aristide and Arpit, Devansh and
                 Draxler, Felix and Lin, Min and Hamprecht, Fred A. and
                 Bengio, Yoshua and Courville, Aaron",
    title     = "{On the Spectral Bias of Neural Networks}",
    booktitle = "International Conference on Machine Learning",
    year      = "2019",
    eprint    = "1806.08734",
    archivePrefix = "arXiv",
    primaryClass  = "stat.ML"
}

@article{Gregory1982,
    author  = "Gregory, J. A. and Delbourgo, R.",
    title   = "{Piecewise Rational Quadratic Interpolation to Monotonic Data}",
    journal = "IMA J. Numer. Anal.",
    volume  = "2",
    number  = "2",
    pages   = "123--130",
    year    = "1982",
    doi     = "10.1093/imanum/2.2.123"
}

@inproceedings{Sitzmann2020siren,
    author    = "Sitzmann, Vincent and Martel, Julien N. P. and Bergman,
                 Alexander W. and Lindell, David B. and Wetzstein, Gordon",
    title     = "{Implicit Neural Representations with Periodic Activation Functions}",
    booktitle = "Advances in Neural Information Processing Systems",
    volume    = "33",
    year      = "2020",
    publisher = "Curran Associates, Inc.",
    eprint    = "2006.09661",
    archivePrefix = "arXiv",
    primaryClass  = "cs.CV"
}

@article{Martone:2012nj,
    author = "Martone, Mario and Zupan, Jure",
    title = "{$B^\pm \to D K^\pm$ with direct CP violation in charm}",
    eprint = "1212.0165",
    archivePrefix = "arXiv",
    primaryClass = "hep-ph",
    doi = "10.1103/PhysRevD.87.034005",
    journal = "Phys. Rev. D",
    volume = "87",
    number = "3",
    pages = "034005",
    year = "2013"
}

@article{Rama:2013voa,
    author = "Rama, Matteo",
    title = "{Effect of D-Dbar mixing in the extraction of gamma with B- -{\ensuremath{>}} D0 K- and B- -{\ensuremath{>}} D0 pi- decays}",
    eprint = "1307.4384",
    archivePrefix = "arXiv",
    primaryClass = "hep-ex",
    doi = "10.1103/PhysRevD.89.014021",
    journal = "Phys. Rev. D",
    volume = "89",
    number = "1",
    pages = "014021",
    year = "2014"
}

@article{BaBar:2005jqu,
    author = "Aubert, Bernard and others",
    editor = "Brenner, R. and de los Heros, C. P. and Rathsman, J.",
    collaboration = "BaBar",
    title = "{An amplitude analysis of the decay $B^\pm \to \pi^\pm \pi^\pm \pi^\mp$}",
    eprint = "hep-ex/0507025",
    archivePrefix = "arXiv",
    reportNumber = "BABAR-PUB-05-028, SLAC-PUB-11332",
    doi = "10.1103/PhysRevD.72.052002",
    journal = "Phys. Rev. D",
    volume = "72",
    pages = "052002",
    year = "2005"
}

@article{BESIII:2025nsp,
    author = "Ablikim, Medina and others",
    collaboration = "BESIII",
    title = "{Updated model-independent measurement of the strong-phase differences between $D^0$ and $\bar{D}^0 \to K^{0}_{S/L}\pi^+\pi^-$ decays}",
    eprint = "2503.22126",
    archivePrefix = "arXiv",
    primaryClass = "hep-ex",
    month = "3",
    year = "2025"
}

@inproceedings{Gilman:2026nys,
    author = "Gilman, Alex",
    title = "{Quantum correlation of neutral charmed mesons at BESIII}",
    booktitle = "{32nd International Symposium on Lepton Photon Interactions at High Energies}: {Lepton-Photon 2025}",
    eprint = "2601.16768",
    archivePrefix = "arXiv",
    primaryClass = "hep-ex",
    month = "1",
    year = "2026"
}

\end{document}